\documentclass[twocolumn,pre,aps,showpacs,showkeys,amsmath]{revtex4}
\usepackage{graphicx}
\usepackage{adjustbox}
\usepackage{multirow}

\begin{document}

\title{Effect of Adult-Born Immature Granule Cells on Pattern Separation in The Hippocampal Dentate Gyrus}
\author{Sang-Yoon Kim}
\email{sykim@icn.re.kr}
\author{Woochang Lim}
\email{wclim@icn.re.kr}
\affiliation{Institute for Computational Neuroscience and Department of Science Education, Daegu National University of Education, Daegu 42411, Korea}

\begin{abstract}
Young immature granule cells (imGCs) appear via adult neurogenesis in the hippocampal dentate gyrus (DG). In comparison to mature GCs (mGCs) (born during development), the imGCs exhibit two competing distinct properties such as high excitability and low excitatory innervation. We develop a spiking neural network for the DG, incorporating the imGCs, and investigate their effect on pattern separation (i.e., a process of transforming similar input patterns into less similar output patterns). We first consider the effect of high excitability. The imGCs become very highly active due to their low firing threshold.
Then, because of high activation, strong pattern correlation occurs, which results in pattern integration (i.e., making association between events).
On the other hand, the mGCs exhibit very sparse firing activity due to strongly increased feedback inhibition (caused by the high activation of the imGCs).
As a result of high sparsity, the pattern separation efficacy (PSE) of the mGCs becomes very high.
Thus, the whole population of GCs becomes a heterogeneous one, composed of a (major) subpopulation of mGCs (i.e., pattern separators) with very low activation degree $D_a^{(m)}$ and a (minor) subpopulation of imGCs (i.e., pattern integrators) with very high activation degree $D_a^{(im)}$. In the whole heterogeneous population, the overall activation degree $D_a^{(w)}$ of all the GCs is a little reduced in comparison to the activation degree $D_a^{(out)}$ in the presence of only mGCs without imGCs. However, no pattern separation occurs, due to heterogeneous sparsity, in contrast to the usual intuitive thought that sparsity could improve PSE. Next, we consider the effect of low excitatory innervation for the imGCs, counteracting the effect of their high excitability.
With decreasing the connection probability of excitatory inputs to the imGCs, $D_a^{(im)}$ decreases so rapidly, and their effect becomes weaker. Then, the feedback inhibition to the mGCs is also decreased, leading to increase in $D_a^{(m)}$ of the mGCs. Accordingly, $D_a^{(w)}$ of the whole GCs also increases.
In this case of low excitatory connectivity, the imGCs perform pattern integration. On the other hand, due to increase in $D_a^{(m)}$, the PSE of the mGCs decreases from a high value to a limit value. In the whole population of all the GCs, when the excitatory connection probability decreases through a threshold, pattern separation starts, the overall PSE increases and approaches that of the mGCs. However, due to heterogeneity caused by the imGCs, the overall PSE becomes deteriorated, in comparison with that in the presence of only mGCs.
\end{abstract}

\pacs{87.19.lj, 87.19.lm, 87.19.lv}
\keywords{Hippocampal dentate gyrus, Adult neurogenesis, Immature granule cells, High excitability, Low excitatory innervation, Pattern separation efficacy}

\maketitle

\section{Introduction}
\label{sec:INT}
The hippocampus, composed of the dentate gyrus (DG) and the subregions CA3 and CA1, plays important roles in memory formation, storage, and retrieval
(e.g., episodic and spatial memory) \cite{Gluck,Squire}. In particular, the subregion CA3 has been considered as an autoassociative network, because of extensive recurrent collateral synapses between the pyramidal cells in the CA3 \cite{Marr,Will,Mc,Rolls1,Rolls2a,Rolls2b,Treves1,Treves2,Treves3,Oreilly}. This autoassociative network operates in both the storage and the recall modes. Storage capacity of the autoassociative network implies the number of distinct patterns that can be stored and accurately recalled. Such storage capacity could be increased if the input patterns into the CA3 are sparse (containing few active elements in each pattern) and orthogonalized (nonoverlapping: active elements in one pattern are unlikely to be active in other patterns). This process of transforming a set of input patterns into sparser and orthogonalized patterns is called pattern separation \cite{Marr,Will,Mc,Rolls1,Rolls2a,Rolls2b,Treves1,Treves2,Treves3,Oreilly,Schmidt,Rolls3,Knier,Myers1,Myers2,Myers3,Scharfman,Yim,Chavlis,Kassab,PS1,PS2,PS3,PS4,PS5,PS6,PS7}.

Here, we are concerned about the DG which is the gateway to the hippocampus. The excitatory granule cells (GCs) in the DG receive excitatory inputs from the entorhinal cortex (EC) via the perforant paths (PPs). As a preprocessor for the CA3, the principal GCs perform pattern separation on the input patterns from the EC
by sparsifying and orthogonalizing them, and provide the pattern-separated outputs to the pyramidal cells in the CA3 through the mossy fibers (MFs) \cite{Treves3,Oreilly,Schmidt,Rolls3,Knier,Myers1,Myers2,Myers3,Scharfman,Yim,Chavlis,Kassab}.
Then, a new pattern may be stored in modified collateral synapses between the pyramidal cells in the CA3.
In this way, pattern separation in the DG could facilitate pattern storage in the CA3.

The whole GCs are grouped into the lamellar clusters \cite{Cluster1,Cluster2,Cluster3,Cluster4}. In each cluster, there exist one inhibitory basket cell (BC)
and one inhibitory HIPP (hilar perforant path-associated) cell, together with excitatory GCs. During pattern separation, the GCs show sparse firing activity via the winner-take-all competition \cite{WTA1,WTA2,WTA3,WTA4,WTA5,WTA6,WTA7,WTA8,WTA9,WTA10,WTA}. Only strongly active GCs survive under the feedback inhibitory inputs from the BC and the HIPP cell. We note that, sparsity (resulting from strong feedback inhibition) has been considered to improve the pattern separation efficacy
\cite{Treves3,Oreilly,Schmidt,Rolls3,Knier,Myers1,Myers2,Myers3,Scharfman,Chavlis,Kassab}.

One of the most distinctive characteristics of the DG is occurrence of adult neurogenesis which results in the generation of new GCs during adulthood.
Altman's pioneering studies in adult rat and cat brains for the adult neurogenesis were done decades ago in the 1960s \cite{NG1,NG2,NG3}.
Since then, adult neurogenesis has been shown to be a robust phenomenon, occurring in most mammals, mainly in the subgranular zone of the DG and the subventricular zone of the lateral ventricles \cite{NG4,NG5,NG6}. The new GCs born in the subgranular zone migrate into the granular layer of the DG.
The whole population of GCs is thus composed of mature GCs (mGCs) born during the development and adult-born immature GCs (imGCs).
In contrast to the mGCs, the young adult-born imGCs are known to have marked properties such as high excitability, weak inhibition, and low excitatory innervation \cite{NG7,NG8,NG9,NG10,NG11}.

In this paper, we develop a spiking neural network for the DG, including both mGCs and imGCs; the fraction of the imGCs is 10 $\%$. In our DG network, high excitability of imGCs is considered, and approximately no inhibition is provided to the imGCs.
We first investigate the effect of adult-born imGCs with high excitability on pattern separation \cite{NG7,NG8,NG9,NG10}. The imGCs show high activation due to lower firing threshold [i.e., their activation degree $D_a^{(im)}$ (= 45 $\%$) becomes very high]. As a result, in the subpopulation of the imGCs, output patterns become highly overlapped (i.e, their Pearson's correlation coefficient is very high). Thus, instead of pattern separation, pattern integration (i.e., making association between events) occurs due to strong pattern correlation. On the other hand, the activation degree $D_a^{(m)}$ (= 1.1 $\%$) of the mGCs becomes very low due to strong feedback inhibition from the inhibitory basket cells (BCs) and HIPP (hilar perforant path-associated) cells (caused by high activation of the imGCs). As a result of high sparsity, the efficacy of pattern separation of the mGCs becomes very high.
In this way, the whole population of GCs is a heterogeneous one, consisting of a (major) subpopulation of mGCs (pattern separators) with very low
$D_a^{(m)}$ and a (minor) subpopulation of imGCs (pattern integrators) with very high $D_a^{(im)}$. In the whole heterogeneous population, the overall activation degree $D_a^{(w)}$ of all the GCs is 5.5 $\%$ [a little less than $D_a^{(out)}$ (= 6 $\%$) in the presence of only mGCs without imGCs]. Although $D_a^{(w)}$ is a little reduced (i.e., sparser firing activity), no pattern separation occurs, due to heterogeneous sparsity, in contrast to the usual intuitive thought that sparsity could improve pattern separation efficacy.

Next, we consider the effect of low excitatory innervation for the imGCs, counteracting the effect of high excitability \cite{NG11}.
In the case of mGCs, they receive excitatory inputs from the entorhinal cortex (EC) via perforant paths (PPs) and from the hilar mossy cells (MCs) with the connection probability $p_c$ (= 20 $\%$). On the other hand, the imGCs receive low excitatory drive from the EC via the PPs and from the MCs with lower connection probability $p_c~(=20~ x~\%)$ ($x:$ synaptic connectivity fraction; $ 0 \leq x \leq 1$).

With decreasing $x$ from 1, $D_a^{(im)}$ of the imGCs decreases so rapidly, and their effect becomes weaker. Then, the feedback inhibition to the mGCs is also decreased, and hence $D_a^{(m)}$ of the mGCs becomes increased. Accordingly, $D_a^{(w)}$ of the whole GCs also increases.
In the whole range of $0 \leq x \leq 1,$ the imGCs are good pattern integrators with strong pattern correlation. On the other hand, due to increase in $D^{(m)}$, the pattern separation efficacy of the mGCs decreases from the high value for $x=1$ to a limit value. In the whole population of all the GCs, due to decreased effect of the imGCs, when $x$ decreases through a threshold, pattern separation starts, and then the overall efficacy of pattern separation increases and approaches that of the mGCs. In the limit case of $x=0$ where all imGCs are silent, the limit efficacy of pattern separation in the whole population is lower than that in the presence of only mGCs (without imGCs), mainly because $D_a^{(w)}$ (= 7.3 $\%$) is larger than $D_a$ (= 6 $\%$) in the absence of imGCs. In this way, due to heterogeneity caused by the imGCs (performing pattern integration), the overall efficacy of pattern separation in the whole heterogeneous population of the GCs becomes deteriorated.

This paper is organized as follows. In Sec.~\ref{sec:DGN}, we describe a spiking neural network for the adult neurogenesis in the hippocampal DG. Then, in the main Sec.~\ref{sec:PS}, we investigate the effect of the adult-born imGCs on pattern separation by varying $x$ (synaptic connectivity fraction). Finally, we give summary and discussion in Sec.~\ref{sec:SUM}.

\section{Spiking Neural Network for The Adult Neurogenesis in The Dentate Gyrus}
\label{sec:DGN}
In this section, we describe our spiking neural network for the adult neurogenesis in the DG.
Based on the anatomical and the physiological properties described in \cite{Myers1,Myers2,Chavlis}, we developed the DG spiking neural networks in the works for the winner-take-all competition \cite{WTA}, the sparsely synchronized rhythm \cite{SSR}, and the pattern separation \cite{PS}. Here, we first refine our prior spiking neural networks to include more synaptic connections with a high degree of anatomical and physiological realism \cite{BN1,BN2},
and then incorporate the young adult-born imGCs to complete structure of our spiking neural network for the adult neurogenesis.

Obviously, our spiking neural network will not capture all the detailed anatomical and physiological complexity of the DG. But, with a limited number of essential elements and synaptic connections in our DG network, effect of the imGCs on the pattern separation could be successfully studied. Hence, our spiking neural network model would build a foundation upon which additional complexity may be added and guide further research.

\begin{figure}
\includegraphics[width=\columnwidth]{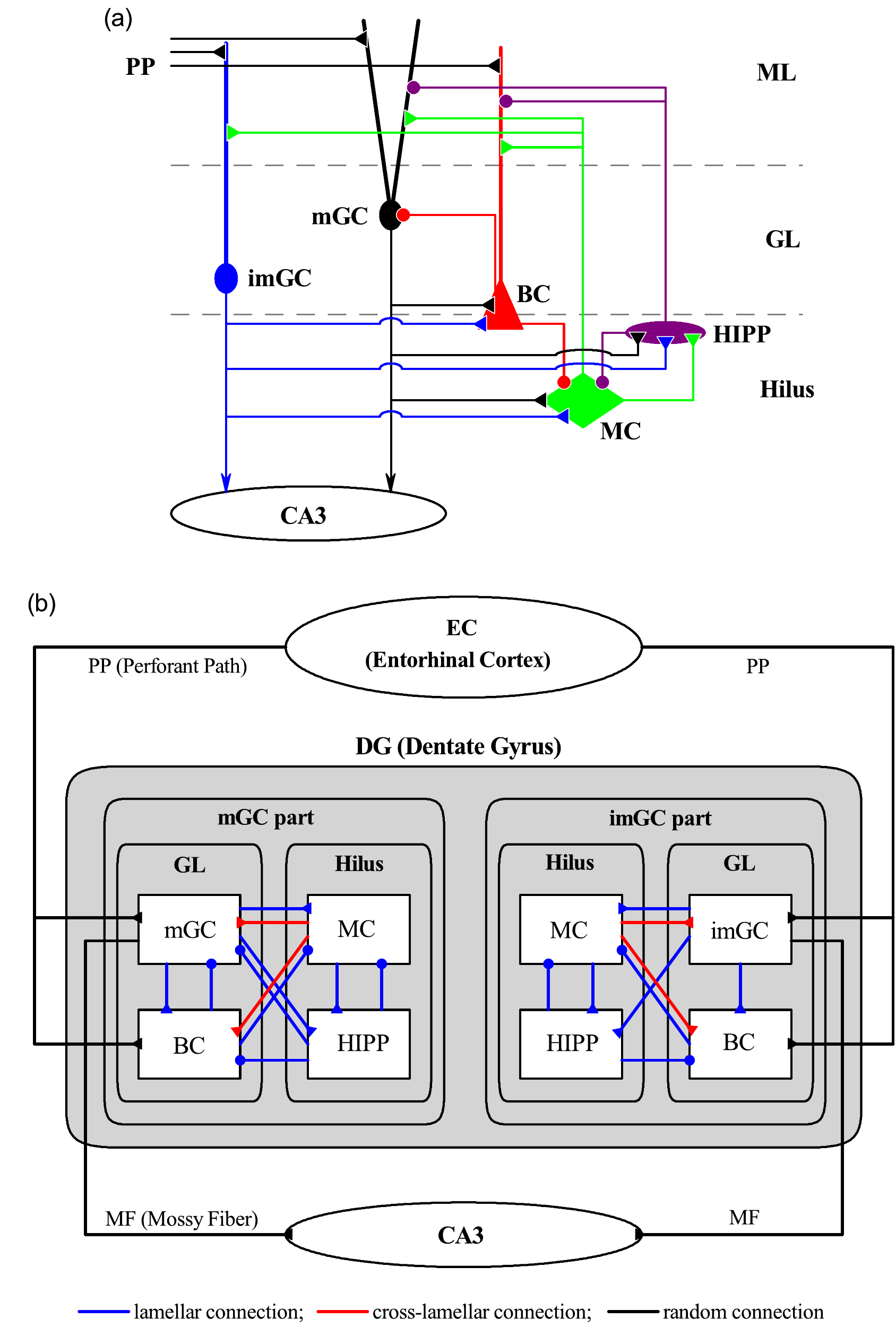}
\caption{Spiking neural network for the hippocampal dentate gyrus (DG). (a) Schematic representation of of major cells and synaptic connections in our DG network incorporating adult-born immature GCs (imGCs). Fraction of the imGCs is 10 $\%$ in the whole population of GCs. Note that there are no inhibitory inputs into the imGCs, in contrast to the case of mGCs. Here, BC, MC, HIPP, PP, GL, and ML represent the basket cell, the mossy cell, the hilar perforant path-associated cell,
perforant path, granular layer, and molecular layer, respectively. (b) Box diagram for our DG network with 3 types of synaptic connections. Blue, red, and black lines represent lamellar, cross-lamellar, and random connections, respectively.
}
\label{fig:DGN}
\end{figure}

\subsection{Architecture of The Spiking Neural Network of The DG}
\label{subsec:SNN}
Figure \ref{fig:DGN} shows (a) schematic representation of major cells and synaptic connections in our DG network incorporating adult-born imGCs and (b) the box diagram for the DG network  with 3 types of lamellar (blue), cross-lamellar (red), and random (black) synaptic connections.
In our DG network, the fraction of imGCs is 10 $\%$ in the whole population of GCs, high excitability of the imGCs is considered, there are no inhibitory inputs into the imGCs, and their low excitatory innervation is also taken into consideration \cite{NG7,NG8,NG9,NG10,NG11}.

In the DG, we consider the granular layer (GL), composed of the excitatory mGCs and imGCs and the inhibitory BCs, and the underlying hilus, consisting of the excitatory MCs and the inhibitory HIPP cells, whose axons project to the upper molecular layer (ML). We note that there are two types of excitatory cells, GCs and MCs, in contrast to the case of the CA3 and CA1 with only one type of excitatory pyramidal cells.

From the outside of the DG, the EC provides the external excitatory inputs randomly to the mGCs, the imGCs, and the inhibitory BCs (with dendrites extending to the outer ML) via PPs \cite{Myers1,Myers2,Myers3,Scharfman,Chavlis}. Thus, both the mGCs and the imGCs receive direct excitatory EC input via PPs (EC $\rightarrow$ mGC and imGCs) through random connections in Fig.~\ref{fig:DGN}(b). The connection probability $p_c$ for EC $\rightarrow$ mGC and BC is 20 $\%$, while $p_c$ for EC $\rightarrow$ imGC is decreased to $20~x$ $\%$ [$x$ (synaptic connectivity fraction); $ 0 \leq x \leq 1 $] due to low excitatory innervation.
Moreover, only the mGCs receive indirect feedforward inhibitory input, mediated by the BCs (EC $\rightarrow$ BC $\rightarrow$ mGC).

In the GL, the whole GCs (i.e., both the mGCs and the imGCs) are grouped into lamellar clusters \cite{Cluster1,Cluster2,Cluster3,Cluster4}, and one inhibitory BC exists in each cluster. Here, the BC (receiving excitation from the whole GCs in the same cluster) provides the feedback inhibition to only all the mGCs via lamellar connections in Fig.~\ref{fig:DGN}(b); a primary mGC-BC feedback loop is formed. Thus, in each cluster the BC provides both the feedforward and the feedback inhibition to all the mGCs in the same cluster.

In the hilus, we also consider lamellar organization for the MCs and HIPP cells \cite{Myers2,Myers3,Scharfman,Hilus6} (i.e., all the MCs and the HIPP cells in the hilus also are grouped into lamellar clusters). As in the case of BC, the HIPP cell receives excitation from the whole GCs in the same cluster, and projects the feedback inhibition to all the mGCs in the same cluster through lamellar connections; a secondary mGC-HIPP feedback loop is formed. Thus, there appear two kinds of feedback loops of mGC-BC and mGC-HIPP.

In our DG network, the MCs play the role of ``controller'' for the activities of the two feedback loops of mGC-BC and mGC-HIPP. Each MC in a cluster receives excitation from the whole GCs in the same cluster (lamellar connection), while it makes excitatory projection randomly to the mGCs and the imGCs in other clusters via cross-lamellar connections \cite{Hilus6}. The connection probability $p_c$ for MC $\rightarrow$ mGC is 20 $\%$, while $p_c$ for MC $\rightarrow$ imGC is decreased to $20~x$ $\%$ ($ 0 \leq x \leq 1 $) because of low excitatory innervation. Thus, the GC-MC driving loop for determining the activities of the controller MCs is formed.

The MCs control the activities of the feedback loops of mGC-BC and mGC-HIPP. Each MC in a cluster receives inhibition from the BC and the HIPP cell in the same cluster (lamellar connection). Then, the MCs in the cluster project excitation to the BCs in other clusters through cross-lamellar connections (the connection probability $p_c$ for MC $\rightarrow$ BC is 20 $\%$) \cite{Hilus6}, while they provide excitation to the HIPP cell in the same cluster (lamellar connection).
Thus, two ``control'' loops of MC-BC and MC-HIPP, controlling the activities of the two feedback loops of mGC-BC and mGC-HIPP, are formed.
Finally, the HIPP cell disinhibits the BC in the same cluster (lamellar connection for HIPP $\rightarrow$ BC); there are no reverse synaptic connections for HIPP $\rightarrow$ BC \cite{BN1,BN2}. Thus, the activity of the BC in a cluster is controlled through excitation from the MCs in other clusters (cross-lamellar connections) and inhibition from the HIPP cell in the same cluster (lamellar connection).

The mGCs in a cluster exhibit sparse firing activity via the winner-take-all competition \cite{WTA1,WTA2,WTA3,WTA4,WTA5,WTA6,WTA7,WTA8,WTA9,WTA10,WTA}. Only strongly active mGCs may survive under the feedback inhibition from the BC and the HIPP cell in the same cluster. Here, the activities of the BC and the HIPP cell are controlled by the controller MCs; in the case of BC, the HIPP cell also disinhibits it. On the other hand, the imGCs receive no inhibition. Particularly, due to their low firing threshold, they become highly active, in contrast to the case of mGCs \cite{NG7,NG8,NG9,NG10}. However, when considering their low excitatory innervation from the EC cells and the MCs, their firing activity is reduced \cite{NG11}.

Based on the anatomical information given in \cite{Myers1,Myers2,Myers3,Scharfman,Chavlis}, we choose the numbers of the GCs, BCs, MCs, and HIPP cells in the DG and the EC cells. As in our prior works \cite{WTA,SSR,PS}, we develop a scaled-down spiking neural network where the total number of excitatory GCs ($N_{\rm GC}$) is 2,000, corresponding to $\frac {1}{500}$ of the $10^6$ GCs found in rats \cite{ANA1}. The fraction of imGCs in the whole population of the GCs is $10~\%$, and hence the number of the imGCs (mGCs) is 200 (1800). The whole GCs (i.e., mGCs and imGCs) are grouped into the $N_c~(=20)$ lamellar clusters \cite{Cluster1,Cluster2,Cluster3,Cluster4}. Then, in each cluster, there are $n_{\rm {GC}}^{(c)}~(=100)$ GCs (i.e., 90 mGC and 10 imGCs) and one inhibitory BC
\cite{Myers2,Myers3,Scharfman}. As a result, the number of the BCs ($N_{\rm BC}$) in the whole DG network becomes 20, corresponding to 1/100 of $N_{\rm GC}$ \cite{GC-BC1,GC-BC2,GC-BC3,GC-BC4,GC-BC5,BN2}.

The EC layer II projects the excitatory inputs to the mGCs, the imGCs, and the BCs via the PPs through random connections \cite{Myers1,Myers2,Myers3,Scharfman,Chavlis}.  The estimated number of the EC layer II cells ($N_{\rm EC}$) is about 200,000 in rats, which corresponds to 20 EC cells per 100 GCs  \cite{ANA3}. Hence, we choose $N_{\rm EC}=400$ in our DG network. Also, the activation degree of the EC cells is chosen as 10$\%$ \cite{ANA4}. Thus, we randomly choose 40 active ones among the 400 EC cells. Each active EC cell is modeled in terms of the Poisson spike train with frequency of 40 Hz \cite{ANA5}.

Next, we consider the hilus, composed of the excitatory MCs and the inhibitory HIPP cells \cite{Hilus1,Hilus2,Hilus3,Hilus4,Hilus5,Hilus6,Hilus7}.
In rats, the number of MCs ($N_{\rm MC}$) is known to change from 30,000 to 50,000, and the estimated number of HIPP cells ($N_{\rm HIPP}$) is about 12,000 \cite{ANA2}. In our scaled-down DG network, we choose $N_{\rm MC}=60$ and $N_{\rm HIPP}=20$. All the MCs and the HIPP cells are also grouped into the 20 lamellar clusters, as in the case of the GCs and the BCs. Hence, in each cluster, there are $n_{\rm {MC}}^{(c)}~(=3)$ MCs and one HIPP cell \cite{Myers2,Myers3,Scharfman}.

With the above information on the numbers of the relevant cells and the synaptic connections between them, we develop a one-dimensional ring network for the
adult neurogenesis in the DG, as in our prior works \cite{WTA,SSR,PS}; e.g., refer to Figs.~1(b1)-1(b3) in \cite{PS} for the schematic diagrams of the ring networks. Due to the ring structure, our spiking neural network has advantage for computational efficiency, and its visual representation may also be easily made.

\subsection{Single Neuron Models and Synaptic Currents in The DG Spiking Neural Network}
\label{subsec:LIF-SC}
As elements of our DG spiking neural network for the adult neurogenesis, we choose leaky integrate-and-fire (LIF) neuron models with additional afterhyperpolarization (AHP) currents which determines refractory periods, as in our prior DG networks \cite{WTA,SSR,PS}.
This LIF neuron model is one of the simplest spiking neuron models \cite{LIF}. Due to its simplicity, it may be easily analyzed and simulated.
It has thus been very popularly used as a spiking neuron model.

The governing equations for evolutions of dynamical states of individual cells in the $X$ population are as follows:
\begin{eqnarray}
C_{X} \frac{dv_{i}^{(X)}(t)}{dt} &=& -I_{L,i}^{(X)}(t) - I_{AHP,i}^{(X)}(t) + I_{ext}^{(X)} - I_{syn,i}^{(X)}(t), \nonumber \\
&& \;\;\; i=1, \cdots, N_{X}, \label{eq:GE}
\end{eqnarray}
where $N_X$ is the total number of cells in the $X$ population, $X=$ mGC, imGC, and BC in the granular layer and $X=$ MC and HIPP in the hilus.
In Eq.~(\ref{eq:GE}), $C_{X}$ (pF) represents the membrane capacitance of the cells in the $X$ population, and the dynamical state of the $i$th cell in the $X$ population at a time $t$ (msec) is characterized by its membrane potential $v_i^{(X)}(t)$ (mV). We note that the time-evolution of $v_i^{(X)}(t)$ is governed by 4 types of currents (pA) into the $i$th cell in the $X$ population; the leakage current $I_{L,i}^{(X)}(t)$, the AHP current $I_{AHP,i}^{(X)}(t)$, the external constant current $I_{ext}^{(X)}$ (independent of $i$), and the synaptic current $I_{syn,i}^{(X)}(t)$.

The equation for a single LIF neuron model (without the AHP current and the synaptic current) describes a simple parallel resistor-capacitor (RC) circuit.
In this case, the 1st type of leakage current is due to the resistor and the integration of the external current is due to the capacitor which is in parallel to the
resistor. When its membrane potential reaches a threshold, a neuron fires a spike, and then the 2nd type of AHP current follows. As the decay time of the AHP current is increased, the refractory period becomes longer. Here, we consider a subthreshold case where the 3rd type of external constant current is zero
(i.e., $I_{ext}^{(X)}=0$) \cite{Chavlis}.

The 1st type of leakage current $I_{L,i}^{(X)}(t)$ for the $i$th cell in the $X$ population is given by:
\begin{equation}
I_{L,i}^{(X)}(t) = g_{L}^{(X)} (v_{i}^{(X)}(t) - V_{L}^{(X)}),
\label{eq:Leakage}
\end{equation}
where $g_L^{(X)}$ and $V_L^{(X)}$ denote conductance (nS) and reversal potential for the leakage current, respectively.
The $i$th cell fires a spike when its membrane potential $v_i^{(X)}$ reaches a threshold $v_{th}^{(X)}$ at a time $t_{f,i}^{(X)}$.
Then, the 2nd type of AHP current $I_{AHP,i}^{(X)}(t)$ follows after spiking (i.e., $t \geq t_{f,i}^{(X)}$), :
\begin{equation}
I_{AHP,i}^{(X)}(t) = g_{AHP}^{(X)}(t) ~(v_{i}^{(X)}(t) - V_{AHP}^{(X)})~~~{\rm ~for~} \; t \ge t_{f,i}^{(X)}.
\label{eq:AHP1}
\end{equation}
Here, $V_{AHP}^{(X)}$ represents the reversal potential for the AHP current, and the conductance $g_{AHP}^{(X)}(t)$ is given by an exponential-decay
function:
\begin{equation}
g_{AHP}^{(X)}(t) = \bar{g}_{AHP}^{(X)}~  e^{-(t-t_{f,i}^{(X)})/\tau_{AHP}^{(X)}},
\label{eq:AHP2}
\end{equation}
where $\bar{g}_{AHP}^{(X)}$ and $\tau_{AHP}^{(X)}$ denote the maximum conductance and the decay time constant for the AHP current, respectively.
With increasing $\tau_{AHP}^{(X)}$, the refractory period becomes longer.

The parameter values of the capacitance $C_X$, the leakage current $I_L^{(X)}(t)$, and the AHP current $I_{AHP}^{(X)}(t)$ are the same as those
in our prior DG networks \cite{WTA,SSR,PS}, and refer to Table 1 in \cite{WTA}. These parameter values are based on physiological properties of the GC, BC, MC,
and HIPP cell \cite{Chavlis,Hilus3}.

\begin{figure}
\includegraphics[width=0.7\columnwidth]{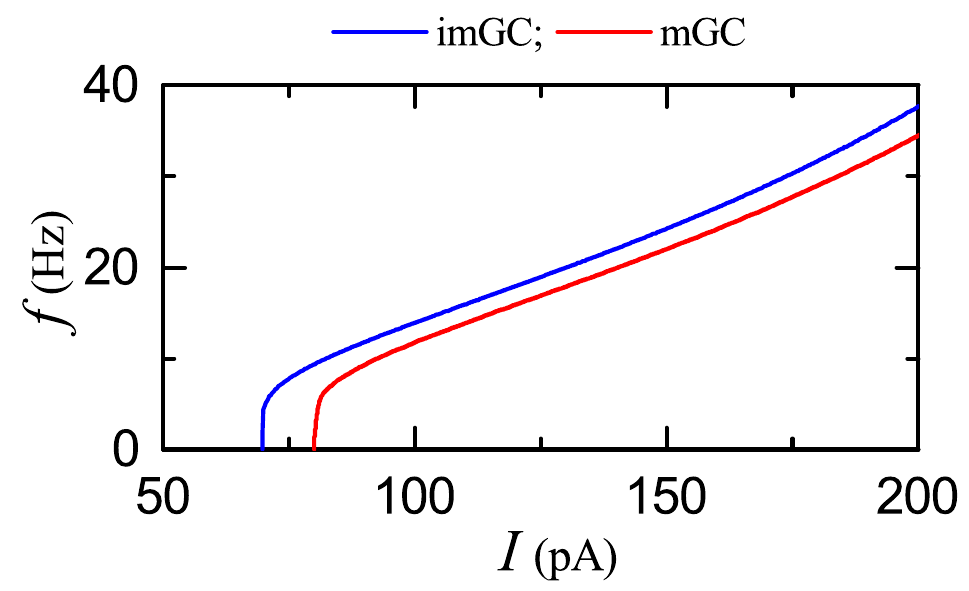}
\caption{Firing transitions of mature GCs (mGCs) and adult-born immature GCs (imGCs). $f-I$ ($f:$ firing rate and $I:$ current) curve for the mature GC (mGC) (red line) and the imGC (blue line).
}
\label{fig:f-I}
\end{figure}

We note that, the GC in Table 1 in \cite{WTA} corresponds to the mGC. The imGCs also have the same parameter values as those of the mGC, except
for the leakage reversal potential $V_L$. The mGC with $V_L=-75$ mV exhibits a spiking transition when passing a threshold $I^*=80$ mV. Here, we consider a case that the imGC has an increased leakage reversal potential of $V_L=-72$ mV, which could lead to intrinsic high excitability. Then, it shows a firing transition when passing $I^*=69.7$ pA. In this way, the imGC may have a lower firing threshold \cite{NG7,NG8,NG9,NG10}, which is well shown
in Fig.~\ref{fig:f-I} for the $f-I$ (i.e., firing rate-current) curves of the mGC (red curve) and the imGC (blue curve).

\begin{table*}
\caption{Parameters for the synaptic currents $I_R^{({\rm GC},S)}(t)$ into the GCs (granule cells). The whole population of the GCs is composed of a major subpopulation of mGCs (mature GCs) and a minor subpopulation of imGCs (immature GCs). Both the mGCs and the imGCs receive the excitatory inputs from the EC (entorhinal cortex) cells and the hilar MCs (mossy cells); synaptic parameters for the excitatory inputs are valid for both the mGCs and the imGCs. In addition,
the mGCs receive the feedforward and feedback inhibitory inputs from the BCs (basket cells) and the feedback inhibitory input from the HIPP (hilar perforant-associated) cells, while there are no inhibitory inputs into the imGCs.
}
\label{tab:Synparm1}
\begin{tabular}{|c|c|c|c|c|c|c|}
\hline
Target Cells ($T$) & \multicolumn{6}{c|}{GC} \\
\hline
Source Cells ($S$) & \multicolumn{2}{c|}{EC} & BC & HIPP & \multicolumn{2}{c|}{MC} \\
\hline
Receptor ($R$) & AMPA & NMDA & GABA & GABA & AMPA & NMDA \\
\hline
$K_{R}^{(T,S)}$ & 0.89 & 0.15 & 15.0 & 3.0 & 0.07 & 0.01 \\
\hline
$\tau_{R,r}^{(T,S)}$ & 0.1 & 0.33 & 0.9 & 0.5 & 0.1 & 0.33 \\
\hline
$\tau_{R,d}^{(T,S)}$ & 2.5 & 50.0 & 6.8 & 6.0& 2.5 & 50.0 \\
\hline
$\tau_{R,l}^{(T,S)}$ & 3.0 & 3.0 & 0.85 & 1.6 & 3.0 & 3.0 \\
\hline
$V_{R}^{(S)}$ & 0.0 & 0.0 & -86.0 & -86.0 & 0.0 & 0.0 \\
\hline
\end{tabular}
\end{table*}

\begin{table*}
\caption{Parameters for the synaptic currents $I_R^{({\rm BC},S)}(t)$ into the BCs (basket cells). The BCs receive the excitatory inputs from the EC (entorhinal cortex) cells, the GCs (granulce cells; both mGCs and imGCs) and the MCs (mossy cells) and the inhibitory input from the HIPP (hilar perforant-associated) cells
}
\label{tab:Synparm2}
\begin{tabular}{|c|c|c|c|c|c|c|c|}
\hline
Target Cells ($T$) & \multicolumn{7}{c|}{BC} \\
\hline
Source Cells ($S$) & \multicolumn{2}{c|}{EC} & \multicolumn{2}{c|}{GC} & \multicolumn{2}{c|}{MC} & HIPP\\
\hline
Receptor ($R$) & AMPA & NMDA & AMPA & NMDA & AMPA & NMDA & GABA \\
\hline
$K_{R}^{(T,S)}$ & 0.75 & 0.13 & 0.38 & 0.02 & 6.14 & 0.36 & 9.22 \\
\hline
$\tau_{R,r}^{(T,S)}$ & 2.0 & 6.6 & 2.5 & 10.0 & 2.5 & 10.0 & 0.4 \\
\hline
$\tau_{R,d}^{(T,S)}$ & 6.3 & 126.0 & 3.5 & 130.0 & 3.5 & 130.0 & 5.8 \\
\hline
$\tau_{R,l}^{(T,S)}$ & 3.0 & 3.0 & 0.8 & 0.8 & 3.0 & 3.0 & 1.6\\
\hline
$V_{R}^{(S)}$ & 0.0 & 0.0 & 0 & 0 & 0.0 & 0.0 & -86.0 \\
\hline
\end{tabular}
\end{table*}

\begin{table*}
\caption{Parameters for the synaptic currents $I_R^{(T,S)}(t)$ into the MCs (mossy cells) and the HIPP (hilar perforant-associated) cells. The MCs receive the excitatory inputs from the GCs (granule cells; both mGCs and imGCs) and the inhibitory inputs from the BCs (basket cells) and the HIPP (hilar perforant-associated) cells. The HIPP cells receive the excitatory inputs from the GCs (both mGCs and imGCs) and the MCs.
}
\label{tab:Synparm3}
\begin{tabular}{|c|c|c|c|c|c|c|c|c|}
\hline
Target Cells ($T$) & \multicolumn{4}{c|}{MC} & \multicolumn{4}{c|}{HIPP cell} \\
\hline
Source Cells ($S$) & \multicolumn{2}{c|}{GC} & BC & HIPP cell & \multicolumn{2}{c|}{GC} & \multicolumn{2}{c|}{MC} \\
\hline
Receptor ($R$) & AMPA & NMDA & GABA & GABA & AMPA & NMDA & AMPA & NMDA  \\
\hline
$K_{R}^{(T,S)}$ & 9.58 & 1.71 & 3.08 & 2.05 & 0.08 & 0.004 & 4.09 & 0.25  \\
\hline
$\tau_{R,r}^{(T,S)}$ & 0.5 & 4.0 & 0.3 & 0.5 & 0.3 & 1.2 & 0.9 & 3.6  \\
\hline
$\tau_{R,d}^{(T,S)}$ & 6.2 & 100.0 & 3.3 & 6.0 & 0.6 & 22.2 & 3.6 & 133.7 \\
\hline
$\tau_{R,l}^{(T,S)}$ & 1.5 & 1.5 & 1.5 & 1.0 & 1.5 & 1.5 & 3.0 & 3.0  \\
\hline
$V_{R}^{(S)}$ & 0.0 & 0.0 & -86.0 & -86.0 & 0.0 & 0.0 & 0.0 & 0.0  \\
\hline
\end{tabular}
\end{table*}

Next, we consider the 4th type of synaptic current $I_{syn,i}^{(X)}(t)$ into the $i$th cell in the $X$ population, composed of the following 3 types of synaptic currents:
\begin{equation}
I_{syn,i}^{(X)}(t) = I_{{\rm AMPA},i}^{(X,Y)}(t) + I_{{\rm NMDA},i}^{(X,Y)}(t) + I_{{\rm GABA},i}^{(X,Z)}(t).
\label{eq:ISyn1}
\end{equation}
Here, $I_{{\rm AMPA},i}^{(X,Y)}(t)$ and $I_{{\rm NMDA},i}^{(X,Y)}(t)$ are the excitatory AMPA ($\alpha$-amino-3-hydroxy-5-methyl-4-isoxazolepropionic acid) receptor-mediated and NMDA ($N$-methyl-$D$-aspartate) receptor-mediated currents from the presynaptic source $Y$ population to the postsynaptic $i$th neuron in the target $X$ population, respectively. In contrast, $I_{{\rm GABA},i}^{(X,Z)}(t)$ is the inhibitory $\rm GABA_A$ ($\gamma$-aminobutyric acid type A) receptor-mediated current from the presynaptic source $Z$ population to the postsynaptic $i$th neuron in the target $X$ population.

Like the case of the AHP current, the $R$ (= AMPA, NMDA, or GABA) receptor-mediated synaptic current $I_{R,i}^{(T,S)}(t)$ from the presynaptic source $S$ population to the $i$th postsynaptic cell in the target $T$ population is given by:
\begin{equation}
I_{R,i}^{(T,S)}(t) = g_{R,i}^{(T,S)}(t)~(v_{i}^{(T)}(t) - V_{R}^{(S)}).
\label{eq:ISyn2}
\end{equation}
Here, $g_{(R,i)}^{(T,S)}(t)$ and $V_R^{(S)}$ represent synaptic conductance and synaptic reversal potential
(determined by the type of the presynaptic source $S$ population), respectively.

In the case of the $R$ (=AMPA and GABA)-mediated synaptic currents, we get the synaptic conductance $g_{R,i}^{(T,S)}(t)$ from:
\begin{equation}
g_{R,i}^{(T,S)}(t) = K_{R}^{(T,S)} \sum_{j=1}^{N_S} w_{ij}^{(T,S)} ~ s_{j}^{(T,S)}(t),
\label{eq:ISyn3}
\end{equation}
where $K_{R}^{(T,S)}$ is the synaptic strength per synapse for the $R$-mediated synaptic current
from the $j$th presynaptic neuron in the source $S$ population to the $i$th postsynaptic cell in the target $T$ population.
The inter-population synaptic connection from the source $S$ population (with $N_s$ cells) to the target $T$ population is given by the connection weight matrix
$W^{(T,S)}$ ($=\{ w_{ij}^{(T,S)} \}$) where $w_{ij}^{(T,S)}=1$ if the $j$th cell in the source $S$ population is presynaptic to the $i$th cell
in the target $T$ population; otherwise $w_{ij}^{(T,S)}=0$. The fraction of open ion channels at time $t$ is also represented by $s^{(T,S)}(t)$.

In contrast, in the NMDA-receptor case, some of the postsynaptic NMDA channels are blocked by the positive magnesium ion ${\rm Mg}^{2+}$
\cite{NMDA}. Hence, the conductance in the case of NMDA receptor is given by \cite{Chavlis}:
\begin{equation}
g_{R,i}^{(T,S)}(t) = {\widetilde K}_R^{(T,S)}~f(v^{(T)}(t))~\sum_{j=1}^{N_S} w_{ij}^{(T,S)} ~ s_{j}^{(T,S)}(t).
\label{eq:NMDA}
\end{equation}
Here, ${\widetilde K}_R^{(T,S)}$ is the synaptic strength per synapse, and the fraction of NMDA channels that are not blocked by the ${\rm Mg}^{2+}$ ion is given by a sigmoidal function $f(v^{(T)}(t))$:
\begin{equation}
f(v^{(T)}(t)) = \frac{1}{1+\eta\cdot [{\rm Mg}^{2+}]_o \cdot \exp(-\gamma \cdot v^{(T)}(t))}.
\end{equation}
Here, $v^{(T)}(t)$ is the membrane potential of the target cell, $[{\rm Mg}^{2+} ]_o$ is the outer ${\rm Mg}^{2+}$ concentration, $\eta$ denotes the sensitivity of ${\rm Mg}^{2+}$ unblock, $\gamma$ represents the steepness of ${\rm Mg}^{2+}$ unblock, and the values of parameters change depending on the target cell \cite{Chavlis}.
For simplicity, some approximation to replace $f(v^{(T)}(t))$ with $\langle f(v^{(T)}(t))\rangle$ [i.e., time-averaged value of $f(v^{(T)}(t))$ in the range of $v^{(T)}(t)$ of the target cell] has been done in \cite{SSR}. Then, an effective synaptic strength $K_{\rm NMDA}^{(T,S)} (={\widetilde K}_{\rm NMDA}^{(T,S)} \langle f(v^{(T)}(t))\rangle$) was introduced by absorbing $\langle f(v^{(T)}(t))\rangle$ into $K_{\rm NMDA}^{(T,S)}$. Thus, with the scaled-down effective synaptic strength $K_{\rm NMDA}^{(T,S)}$ (containing the blockage effect of the ${\rm Mg}^{2+}$ ion), the conductance $g$ for the NMDA receptor may also be well approximated in the same form of conductance as the other AMPA and GABA receptors in Eq.~(\ref{eq:ISyn3}). Thus, we get all the effective synaptic strengths $K_{\rm NMDA}^{(T,S)}$ from the synaptic strengths $\widetilde{K}_{\rm NMDA}^{(T,S)}$ in \cite{Chavlis} by considering the average blockage effect of the ${\rm Mg}^{2+}$ ion.
Consequently, we can use the same form of synaptic conductance of Eq.~(\ref{eq:ISyn3}) in all the cases of $R=$ AMPA, NMDA, and GABA.

The postsynaptic ion channels are opened through binding of neurotransmitters (emitted from the source $S$ population) to receptors in the target
$T$ population. The fraction of open ion channels at time $t$ is represented by $s^{(T,S)}(t)$. The time course of $s_j^{(T,S)}(t)$ of the $j$th cell
in the source $S$ population is given by a sum of double exponential functions $E_{R}^{(T,S)} (t - t_{f}^{(j)}-\tau_{R,l}^{(T,S)})$:
\begin{equation}
s_{j}^{(T,S)}(t) = \sum_{f=1}^{F_{j}^{(s)}} E_{R}^{(T,S)} (t - t_{f}^{(j)}-\tau_{R,l}^{(T,S)}).
\label{eq:ISyn4}
\end{equation}
Here, $t_f^{(j)}$ and $F_j^{(s)}$ are the $f$th spike time and the total number of spikes of the $j$th cell in the source $S$ population, respectively, and
$\tau_{R,l}^{(T,S)}$ is the synaptic latency time constant for $R$-mediated synaptic current.
The double exponential-decay function $E_{R}^{(T,S)} (t)$ (corresponding to contribution of a presynaptic spike occurring at $t=0$ in the absence of synaptic latency)
is given by:
\begin{equation}
E_{R}^{(T,S)}(t) = \frac{1}{\tau_{R,d}^{(T,S)}-\tau_{R,r}^{(T,S)}} \left( e^{-t/\tau_{R,d}^{(T,S)}} - e^{-t/\tau_{R,r}^{(T,S)}} \right) \cdot \Theta(t). \label{eq:ISyn5}
\end{equation}
Here, $\Theta(t)$ is the Heaviside step function: $\Theta(t)=1$ for $t \geq 0$ and 0 for $t <0$, and $\tau_{R,r}^{(T,S)}$ and $\tau_{R,d}^{(T,S)}$ are synaptic rising and decay time constants of the $R$-mediated synaptic current, respectively.

In comparison with our prior DG networks \cite{WTA,SSR,PS}, we include more synaptic connections with a high degree of anatomical and physiological realism \cite{BN1,BN2}, and incorporate the imGCs. Thus, a new feedforward inhibition, mediated by the BCs, is provided to the mGCs, and
there appear two feedback loops of mGC-BC and mGC-HIPP, (projecting feedback inhibition to the mGCs), the activities of which
are controlled by the two control loops of MC-BC and MC-HIPP (MCs: controllers).

Finally, we present the parameter values for the synaptic strength per synapse $K_{R}^{(T,S)}$, the synaptic rising time constant $\tau_{R,r}^{(T,S)}$, synaptic decay time constant $\tau_{R,d}^{(T,S)}$, synaptic latency time constant $\tau_{R,l}^{(T,S)}$, and the synaptic reversal potential  $V_{R}^{(S)}$
for the synaptic currents into the GCs (i.e., both mGCs and imGCs) and the BCs in the GL, in Tables \ref{tab:Synparm1} and \ref{tab:Synparm2}, respectively,
and for the synaptic currents into the MCs and the HIPP cells in Table \ref{tab:Synparm3}.
These parameter values are also based on the physiological properties of the relevant cells
\cite{Chavlis,BN1,BN2, SynParm1,SynParm2,SynParm3,SynParm4,SynParm5,SynParm6,SynParm7,SynParm8}.

All of our source codes for computational works were written in C programming language. Numerical integration of the governing equation for the time-evolution of states of individual spiking neurons is done by employing the 2nd-order Runge-Kutta method with the time step 0.1 msec.

\section{Effect of Immature Granule Cells Born via Adult Neurogenesis on Pattern Separation}
\label{sec:PS}
In this section, we study the effect of adult-born imGCs on pattern separation in our spiking neural network, developed in Sec.~\ref{sec:DGN}.
Due to high excitability, the imGCs become very active, while because of low excitatory innervation, their activation degree is decreased.
We investigate the effects of the two competing properties of the imGCs on the activation degrees and the pattern separation efficacy of the imGCs, the mGCs, and the whole GCs.

\subsection{Characterization of Pattern Separation in The Presence of Only The mGCs without The imGCs}
\label{subsec:mGC}
In this subsection, we first consider the case of presence of only the mGCs (without the imGCs) to present the methods characterizing the pattern separation.
As explained in the subsection \ref{subsec:SNN}, the EC provides external excitatory inputs to the mGCs via PPs [see Fig.~\ref{fig:DGN}(a)]
\citep{Myers1,Myers2,Myers3,Scharfman,Chavlis,WTA,SSR}. We characterize pattern separation between the input patterns of the EC cells and the output patterns of the mGCs via integration of the governing equations (\ref{eq:GE}). In each realization, we have a break stage ($0-300$ msec) (for which the network reaches a stable state), and then a stimulus stage ($300-1,300$ msec) follows; the stimulus period $T_s$ (for which network analysis is done) is 1,000 msec. During the stimulus stage, we get the output firings of the mGCs. For characterization of pattern separation between the input and the output patterns, 30 realizations are made.

The input patterns of the 400 EC cells and the output patterns of the 2,000 mGCs are given in terms of binary representations \cite{Myers1,Chavlis}; active and silent cells are denoted by 1 and 0, respectively. Here, active cells exhibit at least one spike during the stimulus stage. In each realization, we first make a random choice of an input pattern $A^{(in)}$ for the EC cells, and then construct another input patterns $B_i^{(in)}$ ($i=1,\dots,9$) from the base input pattern $A^{(in)}$ with the overlap percentage $P_{OL}$ $=90~$\%$, \dots,$ and $10$ $\%$, respectively, as follows \cite{Myers1,Chavlis}.
Among the active EC cells in the pattern $A^{(in)}$, we randomly choose active cells for the pattern $B^{(in)}$ with the probability $P_{OL}~\%$ (e.g., in the case of $P_{OL} = 60 ~\%$, we randomly choose 24 active EC cells among the 40 active EC cells in the base pattern $A^{(in)}$).
The remaining active EC cells in the pattern $B^{(in)}$ are randomly chosen in the subgroup of silent EC cells in the pattern $A^{(in)}$.

We characterize pattern separation between the input and the output patterns by changing the overlap percentage $P_{OL}$.
For a pair of input ($l=in)$ or output ($l=out$) patterns, $A^{(l)}$ and $B^{(l)}$, their pattern distance $D_p^{(l)}$ is given by \cite{Chavlis,PS}:
\begin{equation}
D_p^{(l)} = {\frac {O^{(l)}} {D_a^{(l)}} }.
\label{eq:PD}
\end{equation}
Here, $D_a^{(l)}$ is the average activation degree of the two patterns $A^{(l)}$ and $B^{(l)}$:
\begin{equation}
D_a^{(l)} = {\frac {(D_a^{(A^{(l)})} + D_a^{(B^{(l)})})} {2} },
\label{eq:AD}
\end{equation}
and $O^{(l)}$ is the orthogonalization degree between $A^{(l)}$ and $B^{(l)}$, denoting their ``dissimilarity'' degree.
Then, as the average activation degree is lower and the orthogonalization degree is higher, the pattern distance between the two patterns $A^{(l)}$ and $B^{(l)}$ increases.

\begin{figure}
\includegraphics[width=\columnwidth]{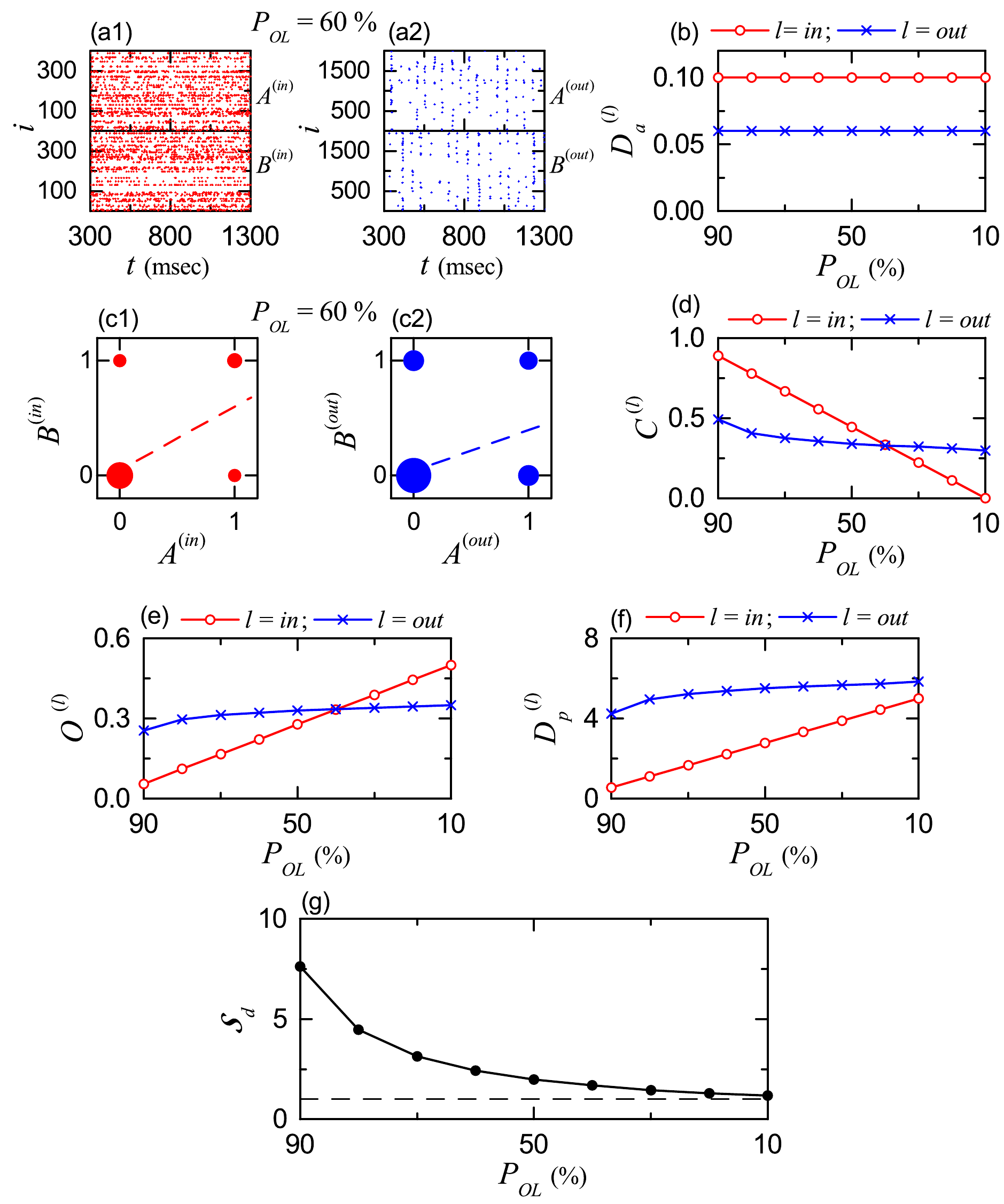}
\caption{Characterization of pattern separation between the input and the output patterns in the presence of only the mGCs without imGCs.
(a1) Raster plots of spikes of ECs for the input patterns $A^{(in)}$ and $B^{(in)}$ in the case of overlap percentage $P_{OL}=60 \%$.
(a2) Raster plots of spikes of GCs for the output patterns $A^{(out)}$ and $B^{(out)}$.
(b) Plots of average activation degree $D_a^{(l)}$ versus $P_{OL}$ for the input ($l=in$; red circle) and the output ($l=out$, blue cross) patterns.
Plots of the diagonal elements (0, 0) and (1, 1) and the anti-diagonal elements (1, 0) and (0, 1) for the spiking activity (1: active; 0: silent)
in the pair of (c1) input ($l=in$) and (c2) output ($l=out$) patterns $A^{(l)}$ and $B^{(l)}$ for $P_{OL}=60\%$;
sizes of solid circles, located at (0,0), (1,1), (1,0), and (0,1), are given by the integer obtained by rounding off the number of $5~ \log_{10} (n_p)$ ($n_p$: number of data at each location), and a dashed linear least-squares fitted line is also given.
Plots of (d) average pattern correlation degree $C^{(l)}$, (e) average orthogonalization degree $O^{(l)}$, (f) pattern distance $D_p^{(l)}$, and (g)
pattern separation degrees ${\cal S}_d$ versus $P_{OL}$ in the case of the input ($l=in$; red circle) and the output ($l=out$, blue cross) patterns.
}
\label{fig:POL}
\end{figure}

Let $\{ a^{(l)}_i \}$ and $\{ b^{(l)}_i \}$ ($i=1,\dots,N_l$) be the binary representations [1 (0) for the active (silent) cell] of the two patterns $A^{(l)}$ and $B^{(l)}$ ($l=in$ or $out$), respectively; $N_{in}=N_{\rm EC}=400$ and $N_{out}=N_{\rm GC}=2,000$.
Then, the Pearson's correlation coefficient $\rho^{(l)}$ between the two patterns $A^{(l)}$ and $B^{(l)}$ is given by
\begin{equation}
\rho^{(l)} = \frac{\sum_{i=1}^{N_l}\Delta a^{(l)}_i \cdot \Delta b^{(l)}_i}{\sqrt{\sum_{i=1}^{N_l}\Delta {a^{(l)}_i}^2}\sqrt{\sum_{i=1}^{N_l}\Delta {b^{(l)}_i}^2}}.
\label{eq:Rho}
\end{equation}
Here, $\Delta a^{(l)}_i = a^{(l)}_i - \langle {a^{(l)}_i} \rangle$, $\Delta b^{(l)}_i = b^{(l)}_i - \langle {b^{(l)}_i} \rangle$, and
$\langle \cdots \rangle$ represents population average over all cells; the range of $\rho^{(l)}$ is [-1, 1]. Then, the pattern correlation degree
$C^{(l)}$, representing the ``similarity'' degree between the two patterns, is given just by their Pearson's correlation coefficient $\rho^{(l)}$:
\begin{equation}
C^{(l)} = \rho^{(l)}.
\label{eq:Corr}
\end{equation}
Then, the orthogonalization degree $O^{(l)},$ denoting the dissimilarity degree between the two patterns, is given by \cite{PS}:
\begin{equation}
O^{(l)} = {\frac {(1-\rho^{(l)})} {2} },
\label{eq:OD}
\end{equation}
where the range of $O^{(l)}$ is [0, 1].

With $D_a^{(l)}$ and $O^{(l)}$, we can obtain the pattern distances of Eq.~(\ref{eq:PD}), $D_p^{(in)}$ and $D_p^{(out)}$, for the input and the output pattern pairs, respectively. Then, the pattern separation degree ${\cal S}_d,$ representing the pattern separation efficacy, is given by the ratio of $D_p^{(out)}$ to $D_p^{(in)}$:
\begin{equation}
{\cal S}_d = {\frac {D_p^{(out)}} {D_p^{(in)}} }.
\label{eq:PSD}
\end{equation}
If ${\cal S}_d > 1$, the output pattern pair of the mGCs is more dissimilar than the input pattern pair of the EC cells, which results in
occurrence of pattern separation. Otherwise (i.e., ${\cal S}_d < 1$), no pattern separation occurs; instead, pattern ``convergence''
(i.e., $D_p^{(out)} < D_p^{(in)}$) takes place.

As a sample example, we consider the case of $P_{OL}=60~\%$. Figure \ref{fig:POL}(a1) shows the raster plots of spikes of 400 EC cells (i.e. a collection of spike trains of individual EC cells) for the input patterns $A^{(in)}$ and $B^{(in)}$ for $P_{OL}=60~\%$. In this case, the activation degree $D_a^{(in)}$ is chosen as 10 $\%$, independently of the input patterns. Figure \ref{fig:POL}(a2) shows the raster plots of spikes of 2,000 mGCs for the output patterns $A^{(out)}$ and $B^{(out)}$. As shown well in the raster plots of spikes, the mGCs show sparser firings than the EC cells. In this case, the average activation degree of Eq.~(\ref{eq:AD}), $D_a^{(out)}$, is 6 $\%$ (which is obtained via 30 realizations). Figure \ref{fig:POL}(b) shows the plot of the average activation degree $D_a^{(l)}$ versus the overlap percentage $P_{OL}$; red circles represent the case of input patterns ($l=in$) and blue crosses denote the case of output patterns ($l=out$). We note that $D_a^{(out)} = 0.06$ (i.e., 6 $\%$), independently of $P_{OL}$. Then, the sparsity ratio, ${\cal R}_s$ ($= D_a^{(in)} / D_a^{(out)}$), becomes 1.667; the firing activity in the output patterns are 1.667 times as sparse as that in the the input patterns.

Figures \ref{fig:POL}(c1) and (c2) show plots of the diagonal elements (0, 0) and (1, 1) and the anti-diagonal elements (1, 0) and (0, 1) for the spiking activity (1: active; 0: silent) in the pair of input ($l=in$) and output ($l=out$) patterns $A^{(l)}$ and $B^{(l)}$ for $P_{OL}=60\%$, respectively.
In each plot, the sizes of solid circles, located at (0,0), (1,1), (1,0), and (0,1), are given by the integer obtained by rounding off the number of  $5~ \log_{10} (n_p)$ ($n_p$: number of data at each location), and a dashed fitted line is also given. In this case, the Pearson's correlation coefficients of Eq.~(\ref{eq:Rho}) (obtained via 30 realizations) for the pairs of the input and the output patterns are $\rho^{(in)}=0.5556$ and $\rho^{(out)}=0.3550$, which correspond to the slopes of the dashed fitted lines. Then, from Eqs.~(\ref{eq:Corr}) and (\ref{eq:OD}), we obtain the average pattern correlation degree $C^{(l)}$ and the average orthogonalization degrees $O^{(l)}$ for the pairs of the input and the output patterns; $C^{(in)}=0.5556,$ $C^{(out)}=0.3550$, $O^{(in)}=0.2222$ and $O^{(out)}=0.3225$.

Figures \ref{fig:POL}(d) and \ref{fig:POL}(e) show plots of the average pattern correlation degree $C^{(l)}$ and the average orthogonalization degree $O^{(l)}$ versus $P_{OL}$ in the case of the input (red circle) and the output (blue cross) patterns, respectively. Obviously, $C^{(l)}$ and $O^{(l)}$ show oppositely-changing tendencies. Hence, it is enough to discuss only the change in $O^{(l)}$. In the case of the pairs of the input patterns, with decreasing $P_{OL}$ from 90 $\%$ to 10 $\%$, $O^{(in)}$ increases linearly from 0.0556 to 0.5. On the other hand, in the case of the pairs of the output patterns, $O^{(out)}$ begins from a much larger value (0.2543), but slowly increases to 0.3507 for $P_{OL}=10$ $\%$ (which is lower than $O^{(in)}$). Thus, the two lines of $O^{(in)}$ and $O^{(out)}$ cross for $P_{OL} \simeq 40$ $\%$. Hence, for $P_{OL} > 40~\%$, $O^{(out)}$ is larger than $O^{(in)}$ (i.e., the pair of output patterns is more dissimilar than the pair of input patterns). In contrast, for $P_{OL} < 40~\%$, $O^{(out)}$ is less than $O^{(in)}$ (i.e., the pair of output patterns becomes less dissimilar than the pair of input patterns).

With the average activation degrees $D_a^{(l)}$ and the average orthogonalization degrees $O^{(l)}$, we can obtain the pattern distances $D_p^{(l)}$
of Eq.~(\ref{eq:PD}) for the pairs of input and output patterns. Figure \ref{fig:POL}(f) shows plots of the pattern distance $D_p^{(l)}$ versus $P_{OL}$
in the case of the input (red circle) and the output (blue cross) patterns. We note that, for all values of $P_{OL}$,
$D_p^{(out)} >  D_p^{(in)}$ (i.e., the pattern distance for the pair of output patterns is larger than that for the pair of input patterns).
However, with decreasing the overlap percentage $P_{OL}$, the difference between $D_p^{(out)}$ and $D_p^{(in)}$ is found to decrease.

Finally, we obtain the pattern separation degree ${\cal S}_d$ of Eq.~(\ref{eq:PSD}) via the ratio of $D_p^{(out)}$ to $D_p^{(in)}$.
Figure \ref{fig:POL}(g) shows plots of the pattern separation degree ${\cal S}_d$ (representing the pattern separation efficacy) versus $P_{OL}$. As $P_{OL}$ is decreased from 90 $\%$ to 10 $\%$, ${\cal S}_d$ is found to decrease from 7.6273 to 1.1691. Hence, for all values of $P_{OL}$, pattern separation occurs because ${\cal S}_d >1$. However, the smaller $P_{OL}$ is, the lower ${\cal S}_d$ becomes.

\subsection{Effect of The Adult-Born imGCs on Pattern Separation}
\label{subsec:imGC}
In this subsection, we consider a population, composed of imGCs and mGCs; the fraction of the imGCs in the whole population is 10 $\%$.
As shown in Fig.~\ref{fig:f-I}, as a result of increased leakage reversal potential $V_L$, the imGC has lower firing threshold than the mGC (i.e., high excitability), which results in high activation of the imGCs \cite{NG7,NG8,NG9,NG10}. We also note that, the imGC has low excitatory innervation, counteracting the high excitability. In the case of the mGCs, the connection probability $p_c$ from the EC cells and the MCs to the mGCs is 20 $\%$, while in the case of the imGCs,
$p_c$ is decreased to $20~x~\%$ [$x$ (synaptic connectivity fraction); $0 \leq x \leq 1$]. Due to low excitatory drive from the EC cells and the MCs, the activation degree of the imGCs becomes reduced.
With decreasing $x$ from 1 to 0, we investigate the effect of high excitability and low excitatory innervation of the imGCs on the pattern separation efficacy.

For a given $x$, we consider 9 pairs of input patterns $(A^{(in)}, B_i^{(in)})$ $(i=1,\dots, 9$) with the overlap percentage $P_{OL}= 90~\%,\dots,$ and $10~ \%$, respectively. All quantities for the input patterns are independent of $x$. The activation degree $D_a^{(in)}$ is 0.1 (10 $\%$), independently of the pairs.
Next, we get the average Pearson's correlation coefficient $\rho^{(in)}$ between the two input patterns in the following way.
We first obtain the realization-averaged Pearson's correlation coefficients $\{ \langle \rho^{(in)}(i) \rangle_r \}$ ($i=1,\dots,9$ corresponds to $P_{OL}=90~\%,\dots,10~\%$, respectively) via 30 realizations; $\langle \cdots \rangle_r$ represents the average over 30 realizations. With decreasing $P_{OL}$ from 90 $\%$ to 10 $\%$, $\langle \rho^{(in)}(i) \rangle_r$ decreases from 0.8889 to 0.0, respectively. As a representative value, we get the average Pearson's correlation coefficient $\rho^{(in)}~(=0.4444),$ corresponding to the mean of $\{ \langle \rho^{(in)}(i) \rangle_r \}$ over all the 9 pairs. Then, from Eqs.~(\ref{eq:Corr}) and (\ref{eq:OD}), we get the average pattern correlation degree $C^{(in)}$ (= 0.4444) and the average orthogonalization degree $O^{(in)}$ (= 0.2778). In this way,  $\rho^{(in)},$ $C^{(in)}$, and $O^{(in)}$
are obtained via double averaging (i.e., averaging over 30 realizations and 9 pairs). Then, the pattern distance $D_p^{(in)}$ of Eq.~(\ref{eq:PD}) between the two input patterns (given by the ratio of the average orthogonalization degree to the average activation degree) becomes 2.778.

\begin{figure}
\includegraphics[width=0.9\columnwidth]{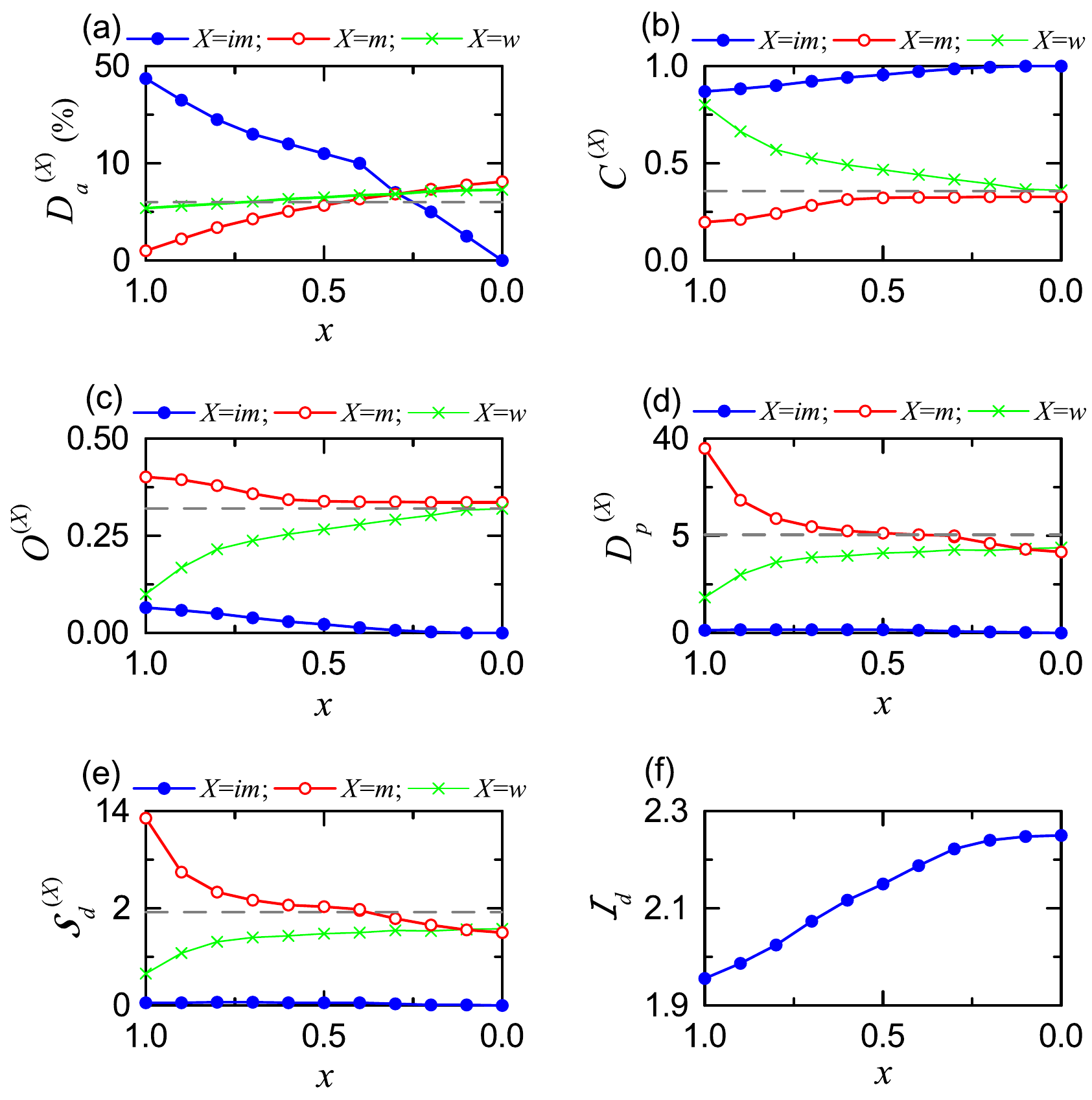}
\caption{Effect of adult-born immature GCs (imGCs) on the pattern separation. (a) Plots of the average activation degree $D_a^{(X)}$ versus $x$ (synaptic connectivity fraction). For clear presentation, we choose two different scales for the vertical axis around $D_a^{(X)}=10$.
(b) Plots of the average pattern correlation degree $C^{(X)}$ versus $x$. (c) Plots of the average orthogonalization degree $O^{(X)}$ versus $x$. (d) Plots of the pattern distance $D_p^{(X)}$ versus $x$. For clear presentation, we choose two different scales for the vertical axis around $D_p^{(X)}=5$. (e) Plots of the pattern separation degree ${\cal S}_d^{(X)}$ versus $x$. For clear presentation, we choose two different scales for the vertical axis around ${\cal S}_d^{(X)}  =2$. In (a)-(e), imGCs ($X=im$), mGCs ($X=m$), and whole GCs ($X=w$) are denoted by blue solid circles, red open circles, and green crosses, respectively. Horizontal dashed lines in (a)-(e) represent $D_a^{(out)}$ (= 6 $\%$), $C^{(out)}$ (= 0.3582), $O^{(out)}$ (= 0.3209), $D_p^{(out)}$ (= 5.3483), and $S_d$ (=1.9252) in the presence of only mGCs (without imGCs), respectively. (f) Plots of the pattern integration degree ${\cal I}_d$ of the imGCs versus $x$.
}
\label{fig:PS}
\end{figure}

As in the above case of input patterns, through double averaging over 30 realizations and 9 pairs, we get the average activation degrees $D_a^{(X)},$ the average Pearson's correlation coefficient $\rho^{(X)},$ the average pattern correlation degree $C^{(X)}$, and the average orthogonalization degrees  ${\cal O}^{(X)}$ in each subpopulation of the imGCs ($X=im$) and the mGCs ($X=m$) and in the whole population ($X=w$).
Figures \ref{fig:PS}(a), \ref{fig:PS}(b), and \ref{fig:PS}(c) show plots of $D_a^{(X)}$, $C^{(X)}$, and ${\cal O}^{(\rm X)}$ versus $x$ [$X=im$ (blue solid circles), $X=m$ (red open circles), and $X=w$ (green crosses)], respectively.
Then, we get the pattern distances $D_p^{(\rm X)}$ of Eq.~(\ref{eq:PD}) (given by the ratio of ${\cal O}^{(X)}$ to
$D_a^{(X)}$), which is shown in Fig.~\ref{fig:PS}(d). Finally, we obtain the pattern separation degree ${\cal S}_d^{(X)}$ of Eq.~(\ref{eq:PSD}) via the ratio of $D_p^{(\rm X)}$ to $D_p^{(in)}$. Figure \ref{fig:PS}(e) shows plots of ${\cal S}_d^{(X)}$ versus $x$. As reference lines, horizontal dashed lines, representing
$D_a^{(out)}$ (= 6 $\%$), $C^{(out)}$ (= 0.3582), $O^{(out)}$ (= 0.3209), $D_p^{(out)}$ (= 5.3483), and $S_d$ (=1.9252) in the presence of only the mGC (without the imGCs) are given  in Figs.~\ref{fig:PS}(a)-\ref{fig:PS}(e), respectively; these values are obtained via averaging over 9 pairs in Fig.~\ref{fig:POL}.

We first consider the case of $x=1$ (where the connection probability $p_c$ from the EC cells and the MCs to the imGCs and the mGCs are the same, 20 $\%$), and discuss the effect of adult-born imGCs with high excitability on pattern separation \cite{NG7,NG8,NG9,NG10}.
The imGCs exhibit high activation due to lower firing threshold [i.e., their average activation degree $D_a^{(\rm im)}$ (= 45 $\%$) becomes very high].
As a result, in the subpopulation of the imGCs, output patterns become highly overlapped (i.e, their average Pearson's correlation coefficient is very high),
which leads to very high average pattern correlation degree $C^{(im)}~(=~0.8692)$ and very low average orthogonalization degree ${\cal O}^{(\rm im)}~(=~0.0654)$. Then, their pattern distance $D_p^{(\rm im)}$ (= 0.145), given by the ratio of ${\cal O}^{(\rm im)}$ to $D_a^{(\rm im)}$, also becomes very low. Consequently, the pattern separation degree ${\cal S}_d^{(\rm im)}$, given by the ratio of $D_p^{(\rm im)}$ to $D_p^{(in)}$, is 0.052.
Since ${\cal S}_d^{(\rm im)} < 1$, no pattern separation occurs, due to their high excitability. On the other hand,
the efficacy of pattern integration (i.e., making association between events) is very high due to high pattern correlation degree $C^{(im)}$.
We introduce the pattern integration degree ${\cal I}_d$ of the imGCs, given by the ratio of the average pattern correlation degree $C^{(im)}$ to the average pattern correlation degree $C^{(in)}$ for the input patterns:
\begin{equation}
   {\cal I}_d = \frac {C^{(im)}} {C^{(in)}},
\label{eq:ID}
\end{equation}
which is in contrast to the pattern separation degree ${\cal S}_d$ of Eq.~(\ref{eq:PSD}).
For $x=1$ the pattern integration degree of the imGCs is high (i.e., ${\cal I}_d=1.9559$).
Figure \ref{fig:PS}(f) shows plots of ${\cal I}_d$ versus $x$ for the imGCs.
With decreasing $x$ from 1 to 0, ${\cal I}_d$ is increased from 1.9559 to 2.2502, because $C^{(im)}$ increases from 0.8692 to 1.
In the whole range of $0 \leq x \leq 1$, the imGCs are good pattern integrators with ${\cal I}_d > 1$.

In contrast, for $x=1$ the mGCs exhibit very sparse firing activity (i.e., their average activation degree $D_a^{(m)}$ (= 1.1 $\%$) of the mGCs becomes very low)
due to strong feedback inhibition from the BCs and the HIPP cells (caused by the high activation of the imGCs). As a result of high sparsity, the
average Pearson's correlation coefficient between the output-pattern pairs becomes very low, which leads to high average orthogonalization degree ${\cal O}^{(m)}$ (= 0.4016). Then, their pattern distance $P_d^{(m)}$ (=36.509) becomes very high. Accordingly, the pattern separation degree ${\cal S}_d^{(m)}$ is 13.142.
Thus, the pattern separation efficacy of the mGCs becomes very high (i.e., the mGCs become good pattern separators), due to high sparsity.

In the above way, the whole population of all the GCs for $x=1$ is a heterogeneous one, composed of a (major) subpopulation of sparsely active mGCs (good pattern separators) with very low $D_a^{(m)}$ and a (minor) subpopulation of highly active imGCs (good pattern integrators) with very high $D_a^{(im)}$; most of active cells congregate in the subpopulation of the imGCs. In the whole heterogeneous population, the overall activation degree $D_a^{(w)}$ of all the GCs is 0.055 (5.5 $\%$) which is a little less than $D_a^{(out)}$ (= 6 $\%$) in the presence of only mGCs (without imGCs). Although $D_a^{(w)}$ is a little decreased (i.e., sparser firing activity), the average Pearson's correlation coefficient between the output-pattern pairs becomes high, due to presence of strongly-correlated imGCs, which leads to
low orthogonalization degree ${\cal O}^{(w)}$ (=0.1004); ${\cal O}^{(w)}$ is also much less than the average orthogonalization degree ${\cal O}^{(out)}$ (= 0.3209)
in the presence of only mGCs. Then, we get the pattern distance $D_p^{(w)}$ (= 1.825) which is also less than $D_p^{(in)}$ (= 2.778). Consequently, the pattern separation degree ${\cal S}_d^{(w)}$ becomes 0.657. Since ${\cal S}_d^{(w)} < 1$, no pattern separation occurs in the whole heterogeneous population for $x=1$, due to heterogeneous sparsity, in contrast to the usual intuitive thought that sparsity could improve pattern separation efficacy; such intuitive thought might be applied only to homogeneous sparsity. Instead of pattern separation, pattern ``convergence'' with $D_p^{(w)} < D_p^{(in)}$ occurs for $x=1$ in the whole heterogeneous population of all the GCs.

Next, with decreasing $x$ from 1, we consider the effect of low excitatory innervation for the imGCs, counteracting the effect of high excitability \cite{NG11}.
In the case of mGCs, they receive excitatory inputs from the EC via PPs and from the hilar MCs with the connection probability $p_c$ (= 20 $\%$). On the other hand, the imGCs receive low excitatory drive via the PPs and from the MCs with lower connection probability $p_c~(=20~ x~\%)$ ($x:$ synaptic connectivity fraction; $ 0 \leq x \leq 1$). As $x$ is decreased from 1, $D_a^{(im)}$ of the imGCs decreases so rapidly, and their effect becomes weaker. Then, the feedback inhibition to the mGCs is also decreased, and hence $D_a^{(m)}$ of the mGCs becomes increased. Accordingly, $D_a^{(w)}$ of the whole GCs also increases.
In the whole range of $0 \leq x \leq 1$, the average pattern correlation degree $C^{(im)}$ of the imGCs are very high, and hence they become good pattern
integrators with the pattern integration degree ${\cal I}_d > 1$ [see Fig.~\ref{fig:PS}(f)].
On the other hand, due to increase in $D^{(m)}$, the pattern separation efficacy of the mGCs decreases from the high value (${\cal S}_d^{(m)}=$ 13.142) for $x=1$ to a limit value (${\cal S}_d^{(m)}=$ 1.495) for $x=0$. In the whole population of all the GCs, due to decreased effect of the imGCs, when $x$ decreases through a threshold $x^*~(=0.92)$, pattern separation (with ${\cal S}_d^{(w)} > 1$) starts, and then the overall pattern separation degree ${\cal S}_d^{(w)}$ increases and approaches a limit value (${\cal S}_d^{(w)}=$ 1.577) for $x=0$ which is a little larger than the limit value of the mGCs.
In the limit case of $x=0$ where all imGCs are silent, the limit pattern separation degree (${\cal S}_d^{(w)}~=~1.577$) in the whole population is lower than that
(${\cal S}_d~=~1.9252$) in the presence of only mGCs (without imGCs), mainly because $D_a^{(w)}$ (= 7.3 $\%$) is larger than $D_a^{(out)}$ (= 6 $\%$) in the absence of imGCs. In this way, due to heterogeneity caused by the imGCs (performing pattern integration), the overall efficacy of pattern separation in the whole heterogeneous population of all the GCs becomes deteriorated.

\section{Summary and Discussion}
\label{sec:SUM}
We investigated the effect of the adult-born imGCs on the pattern separation in a spiking neural network, composed of both mGCs (born during development) and imGCs.  In contrast to the mGCs, the imGCs exhibit two competing distinct properties of high excitability (causing high activation) and low excitatory innervation (reducing activation degree). We first considered the effect of high excitability. The activation degree $D_a^{(im)}$ (= 45 $\%$) of the imGCs was found to be very high due to
lower firing threshold. In this case, the pattern correlation degree $C^{(im)}$ (= 0.8692) also became high, because the outputs were highly overlapped.
Consequently, the imGCs were found to become good pattern integrators (i.e., making association between events) with the pattern integration degree
${\cal I}_d~(= 1.9559)$. In contrast, the activation degree $D_a^{(m)}$ (= 1.1 $\%$) of the mGCs was found to be very low due to strong feedback inhibition from the inhibitory BCs and HIPP cells (caused by high activation of the imGCs). Due to high sparsity, the efficacy of pattern separation of the mGCs  became very high. Thus, the mGCs were found to become good pattern separators with the pattern separation degree ${\cal S}_d~(= 13.142)$.

In the above way, the whole population of all the GCs became a heterogeneous one, composed of a (major) subpopulation of mGCs (good pattern separators) with very low
$D_a^{(m)}$ and a (minor) subpopulation of imGCs (good pattern integrators) with very high $D_a^{(im)}$; most of active cells congregated in the subpopulation of imGCs. In the whole heterogeneous population, the overall activation degree $D_a^{(w)}$ (= 5.5 $\%$) of all the GCs was found to be a little less than $D_a^{(out)}$ (= 6 $\%$) in the presence of only mGCs (without imGCs). However, in spite of sparser firing activity, no pattern separation occurred, because of heterogeneous sparsity, in contrast to the usual intuitive thought that sparsity could improve pattern separation efficacy; such intuitive thought might be applied only to the case of homogeneous sparsity. Instead, pattern convergence with ${\cal S}_d~(= 0.657)$ was found to occur because $D_p^{(w)} < D_p^{(in)}$.

Next, we studied the effect of low excitatory innervation of the imGCs, counteracting the effect of their high excitability; the connection probability $p_c$ from the EC cells and the MCs to the imGCs is $20~x~\%$ [$x$ (synaptic connectivity fraction); $0 \leq x \leq 1$].
As $x$ was decreased from 1 to 0, $D_a^{(im)}$ of the imGCs was found to decrease so rapidly, and hence their effect became weaker.
In contrast to the case of the imGCs, $D_a^{(m)}$ of the mGCs became increased due to decrease in the feedback inhibition from the BCs and the HIPP cells.
Consequently, $D_a^{(w)}$ of the whole GCs also increased. In the whole range of $0 \leq x \leq 1$, the imGCs were found to have high pattern correlation
degree ($0.8692 \leq C^{(im)} \leq 1.0$), and hence they became good pattern integrators with the pattern integration degree ($1.9559 \leq {\cal I}_d \leq 2.2502$).

On the other hand, due to increase in $D_a^{(m)}$, the pattern separation degree ${\cal S}_d^{(m)}$ of the mGCs was found to decrease from the high value
(13.142) for $x=1$ to a limit value (1.495) at $x=0$. Thus, in the whole range of $0 \leq x \leq 1$, the mGCs performed pattern separation with ${\cal S}_d^{(m)} >1$. In the whole population of all the GCs, when $x$ decreases through a threshold $x^*~(=0.92)$, pattern separation (with ${\cal S}_d^{(w)} >1$) was found to start, and then the overall pattern separation degree ${\cal S}_d^{(w)}$ increased and approached a limit (1.577) which was a little larger than the limit (1.495) of the mGCs. However, ${\cal S}_d^{(w)}$ was found to be less than ${\cal S}_d$ (= 1.9252) in the presence of only mGCs (without imGCs).
Thus, due to heterogeneity caused by the imGCs, the pattern separation efficacy in the heterogeneous population became deteriorated, in comparison with that in the presence of only mGCs.

\begin{figure}
\includegraphics[width=\columnwidth]{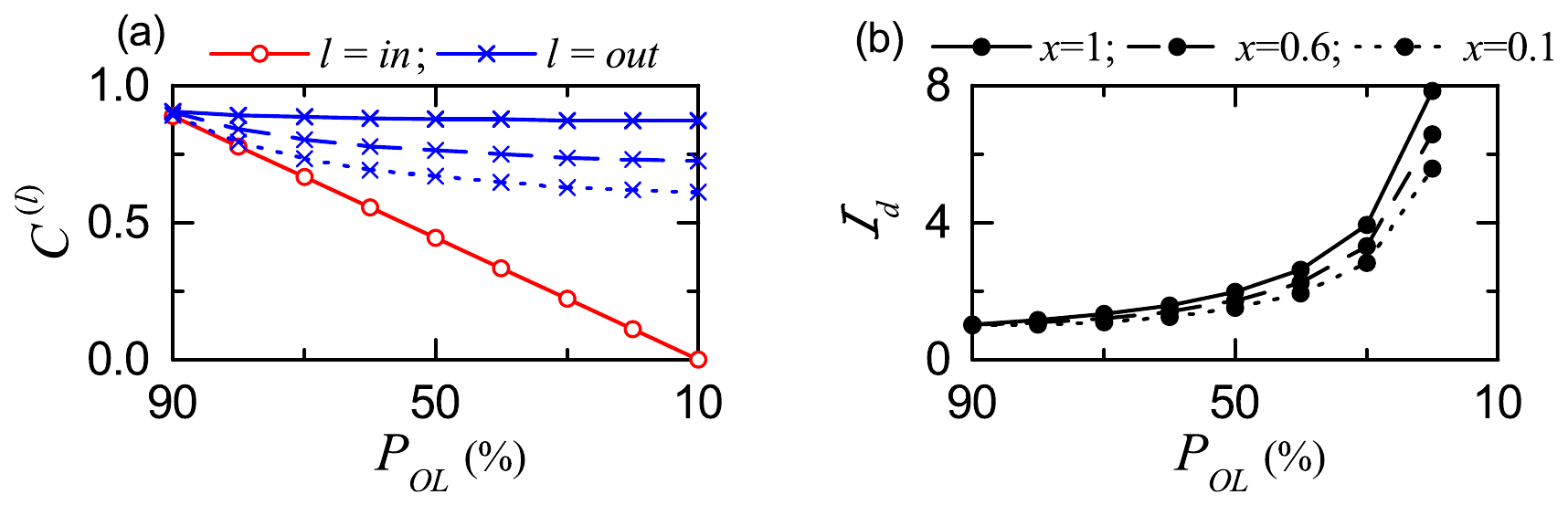}
\caption{Pattern integration in the presence of only imGCs.
(a) Plots of pattern correlation degrees $C^{(l)}$ versus $P_{OL}$; $l=in$ (red) and $l=out$ (blue).
(b) Plots of integration degree ${\cal I}_d$ versus $P_{OL}$; for $P_{OL} = 10~\%$ ${\cal I}_d$ becomes infinity (not shown) because
$C^{(in)}=0$. In (a) and (b), the solid, dashed , and dotted lines correspond to the cases of $x=1$, 0.6, and 0.1, respectively.
}
\label{fig:PI}
\end{figure}

In Fig.~\ref{fig:POL}, we characterized pattern separation by varying the overlap percentage $P_{OL}$ in the homogeneous population of only the mGCs (without
the imGCs). Thus, the mGCs were found to perform pattern separation. It was also found that, the smaller $P_{OL}$ is, the lower the pattern separation degree
${\cal S}_d$ becomes (i.e., the pattern separation efficacy becomes better for similar input patterns, while in the case of dissimilar input patterns, the pattern separation efficacy becomes worse). For comparison, we consider another homogeneous population of only the imGCs (without the mGCs) to more clearly understand the role of the imGCs. Figure \ref{fig:PI}(a) shows the plots of the pattern correlation degree ${\cal C}^{(l)}$ versus $P_{OL}$ for the pair of input patterns
[$l=in$ (red)] and output patterns [$l=out$ (blue)]; in the case of $l=out$, the solid, dashed, and dotted lines correspond to the cases of $x=1$, 0.6, and 0.1, respectively. Then, the pattern integration degree ${\cal I}_d$ is given by the ratio of ${\cal C}^{(out)}$ to ${\cal C}^{(in)}$.
Figure \ref{fig:PI}(b) shows ${\cal I}_d$ versus $P_{OL}$. [We note that in the case of $P_{OL} = 10~\%$, ${\cal C}^{(in)} =0,$ and hence
${\cal I}_d$ becomes infinity (not shown).] We note that, as $P_{OL}$ is decreased, ${\cal I}_d$ becomes increased, in contrast to the case of ${\cal S}_d$ in Fig.~\ref{fig:POL}. Thus, the pattern integration efficacy becomes better for dissimilar input patterns. Also, as $x$ is decreased from 1, the effect of imGCs becomes weaker, leading to decrease in ${\cal I}_d$.

As discussed above, the pattern separation efficacy in the heterogeneous population of all the GCs (composed of both mGCs and imGCs) was found to get deteriorated, due to presence of the imGCs (good pattern integrators). However, we note that the pattern separation may not always be a strict requirement for accurate neural encoding. In the homogeneous population of only the mGCs (without the imGCs), memory storage capacity (representing the number of distinct patterns which may be stored and accurately recalled) could be increased with pattern separation efficacy (facilitating the pattern storage and retrieval) \cite{Myers1}. In contrast, in a heterogeneous population of mGCs (pattern separators) and imGCs (pattern integrators), the memory storage capacity might be optimally maximized via mixed encoding
through pattern separation on similar input patterns and pattern integration on very dissimilar input patterns \cite{NG9,Hetero}.
Thus, through mixed encoding, memory resolution (corresponding to the extent of information incorporated into memories) could be increased, which would result in reduction in memory interference. In this way, the imGCs (good integrators for very dissimilar input patterns) could make contribution to increase in memory storage capacity, although they have tendency to reduce the pattern separation efficacy. Through cooperation of pattern separation for similar input patterns and
pattern integration for very dissimilar input patterns, the heterogeneous population of the mGCs and the imGCs might achieve superior pattern encoding than
the homogeneous population of only the mGCs (performing purely sparse coding). This speculation on increase in memory resolution via mixed encoding (through cooperation of pattern separation and pattern integration) must be examined in future works.

Finally, we discuss future works. During the pattern separation, sparsely synchronized rhythms appear in the whole population of all the GCs and in each subpopulation of the imGCs and the mGCs. Hence, it would be worthwhile to investigate their population and individual firing behaviors and to discuss their quantitative relationship with the pattern separation efficacy. As in \cite{SSR,PS}, population and individual firing behaviors in the sparsely synchronized rhythms in the subpopulations of the imGCs ($X=im$), the mGCs ($X=m$) and in the whole population ($X=w$) may be characterized in terms of the amplitude measure ${\cal M}_a^{(X)}$ (representing the popupation synchronization degree) \cite{AM} and the coefficient of variation $CV^{(X)}$ (characterizing the irregularity degree of individual single-cell discharges) \cite{CV}, respectively. Then, we could investigate the quantitative relationship between ${\cal M}_a^{(X)}$ and $CV^{(X)}$ of the sparsely synchronized rhythms and the pattern separation degree ${\cal S}_d^{(X)}$ (representing the pattern separation efficacy). Next, we also note that the pyramidal cells in the CA3 provide backprojections to the GCs via polysynaptic connections \citep{Myers2,Myers3,Scharfman}. For example, the pyramidal cells send disynaptic inhibition to the mGCs, mediated by the BCs and the HIPP cells in the DG, and they provide trisynaptic inputs to the mGCs, mediated by the MCs (pyramidal cells $\rightarrow$ MC $\rightarrow$ BC or HIPP $\rightarrow$ mGC). These inhibitory backprojections may decrease the activation degree of the mGCs, leading to improvement of pattern separation in the subpopulation of the mGCs. Hence, in future work, it would be meaningful to take into consideration the backprojection for the study of pattern separation in the combined DG-CA3 network. Moreover, in the DG-CA3 network, we could examine the memory storage capacity by getting correct response percentage for a partial or noisy version of cue input patterns in the homogeneous population of only the mGCs and in a heterogeneous population of the mGCs and the imGCs \cite{Myers2}. Then, we could determine which one of the purely sparse encoding (homogeneous case) and the mixed encoding (heterogeneous case) would be superior.

\section*{Acknowledgments}
This research was supported by the Basic Science Research Program through the National Research Foundation of Korea (NRF) funded by the Ministry of Education (Grant No. 20162007688).


\begin{thebibliography}{}
\bibitem{Gluck} M. A. Gluck and C. E. Myers, {\it Gateway to Memory: An Introduction to Neural Network Modeling of the Hippocampus in Learning and
Memory} (MIT Press, Cambridge, 2001).
\bibitem{Squire} L. Squire, {\it Memory and Brain} (Oxford University Press, New York, 1987).
\bibitem{Marr} D. Marr, Phil. Trans. R. Soc. Lond. B {\bf 262}, 23 (1971).
\bibitem{Will} D. Willshaw and J. Buckingham, Phil. Trans. R. Soc, Lond. B{\bf 329}, 205 (1990).
\bibitem{Mc} B. McNaughton and R. Morris, Trends Neurosci. {\bf 10}, 408 (1987).
\bibitem{Rolls1} E. T. Rolls, ``Functions of neuronal networks in the hippocampus and neocortex in memory,'' in
J. H. Byrne and W. O. Berry (eds.), {\it Neural Models of Plasticity: Experimental and Theoretical Approaches}
(Academic Press, San Diego, 1989) pp. 240–265.
\bibitem{Rolls2a} E. T. Rolls, ``The representation and storage of information in neural networks
        in the primate cerebral cortex and hippocampus,'' in R. Durbin, C. Miall, and G.
        Mitchison (eds.), {\it The Computing Neuron} (Addition-Wes;ey, Wokingham, 1989) pp. 125–159.
\bibitem{Rolls2b} E. T. Rolls, ``Functions of neuronal networks in the hippocampus and cerebral cortex in memory,'' in
R. Cotterill (ed.) {\it Models of Brain Function} (Cambridge University Press, New York, 1989) pp. 15 – 33.
\bibitem{Treves1} A. Treves and E. T. Rolls, Network {\bf 2}, 371 (1991).
\bibitem{Treves2} A. Treves and E. T. Rolls, Hippocampus {\bf 2}, 189 (1992).
\bibitem{Treves3} A. Treves and E. T. Rolls, Hippocampus {\bf 4}, 374 (1994).
\bibitem{Oreilly} R. C. O'Reilly and J. C. McClelland, Hippocampus {\bf 4}, 661 (1994).
\bibitem{Schmidt} B. Schmidt, D. F. Marrone, and E. J. Markus, Behav. Brain Res. {\bf 226}, 56 (2012).
\bibitem{Rolls3} E. T. Rolls, Neurobiol. Learn. Mem. {\bf 129}, 4 (2016).
\bibitem{Knier} J. J. Knierim and J. P. Neunuebel, Neurobiol. Learn. Mem. {\bf 129}, 38 (2016).
\bibitem{Myers1} C. E. Myers and H. E. Scharfman, Hippocampus {\bf 19}, 321 (2009).
\bibitem{Myers2} C. E. Myers and H. E. Scharfman, Hippocampus {\bf 21}, 1190 (2011).
\bibitem{Myers3} C. E. Myers, K. Bermudez-Hernandez, and H. E. Scharfman. PLoS ONE {\bf 8}, e68208 (2013).
\bibitem{Scharfman} H. E. Scharfman and C. E. Myers, Neurobiol. Learn. Mem. {\bf 129}, 69 (2016).
\bibitem{Yim} M. Y. Yim, A. Hanuschkin, and J. Wolfart, Hippocampus {\bf 25}, 297 (2015).
\bibitem{Chavlis} S. Chavlis, P. C. Petrantonakis, and P. Poirazi, Hippocampus {\bf 27}, 89 (2017).
\bibitem{Kassab} R. Kassab and F. Alexandre, Brain Struct. Funct. {\bf 223}, 2785 (2018).

\bibitem{PS1} H. Beck, I. V. Goussakov, A. Lie, C. Helmstaedter, and C. E. Elger, J. Neurosci. {\bf 20}, 7080 (2000).
\bibitem{PS2} D. Nitz and B. McNaughton, J. Neurophysiol. {\bf 91}, 863, (2004).
\bibitem{PS3} J. K. Leutgeb, S. Leutgeb, M.-B. Moser, and E. I. Moser, Science {\bf 315}, 961 (2007).
\bibitem{PS4} A. Bakker, C. B. Kirwan, M. Miller, and C. E. L. Stark, Science {\bf 319}, 1640 (2008).
\bibitem{PS5} M. A. Yassa and C. E. L. Stark, Trends Neurosci. {\bf 34}, 515 (2011).
\bibitem{PS6} A. Santoro, Front. Behav. Neurosci. {\bf 7}, 96 (2013).
\bibitem{PS7} M. T. van Dijk and A. A. Fenton, Neuron {\bf 98}, (2018).

\bibitem{Cluster1} P. Andersen, T. V. P. Bliss, and K. K. Skrede, Exp. Brain Res. {\bf 13}, 222 (1971).
\bibitem{Cluster2} D. G Amaral and M. P. Witter, Neurosci. {\bf 31}, 571 (1989).
\bibitem{Cluster3} P. Andersen, A. F. Soleng, and M. Raastad, Brain Res. {\bf 886}, 165 (2000).
\bibitem{Cluster4} R. S. Sloviter and T. L{\o}mo, Front. Neural Circ. {\bf 6}, 102 (2012)

\bibitem{WTA1} R. Coultrip, R. Granger, and G. Lynch, Neural Netw. {\bf 5}, 47 (1992).
\bibitem{WTA2} L. de Almeida, M. Idiart, and J. E. Lisman, J. Neurosci. {\bf 29}, 7497 (2009).
\bibitem{WTA3} P. C. Petrantonakis and P. Poirazi,  Front. Syst. Neurosci. {\bf 8}, 141 (2014).
\bibitem{WTA4} P. C. Petrantonakis and P. Poirazi, PLoS One {\bf 10}, e0117023 (2015).
\bibitem{WTA5} C. Houghton, Behav. Brain Res. {\bf 39}, 28 (2017).
\bibitem{WTA6} C. Espinoza, S. J. Guzman, X. Zhang, and P. Jonas, Nat. Commun. {\bf 9}, 4605 (2018).
\bibitem{WTA7} L. Su, C.-J. Chang, and N. Lynch, Neural Comput. {\bf 31}, 2523 (2019).
\bibitem{WTA8} V. J. Barranca, H. Huang, and G. Kawakita, J. Comput. Neurosci. {\bf 46}, 145 (2019).
\bibitem{WTA9} N. Z. Bielczyk, K. Piskała, M. Płomecka, P. Radzi\'{n}ski, L. Todorova, and U. Fory\'{s}, PLoS One {\bf 14}, e0211885 (2019).
\bibitem{WTA10} Y. Wang, X. Zhang, Q. Xin, W. Hung, J. Florman, J. Huo, T. Xu, Y. Xie, M. J. Alkema, M. Zhen, and Q. Wen,
eLife {\bf 9}, e56942 (2020).

\bibitem{WTA} S.-Y. Kim and W. Lim, Phys. Rev. E {\bf 105}, 014418 (2022).

\bibitem{NG1} J. Altman, Science {\bf 135}, 1127 (1862),
\bibitem{NG2} J. Altman, Anat. Rec. {\bf 145},  573 (1963).
\bibitem{NG3} J. Altman and G. D. Das, J. Comp. Neurol. {\bf 124}, 319 (1965).
\bibitem{NG4} S. A. Bayer, J. Comp. Neurol. {\bf 524}, 2933 (2016).
\bibitem{NG5} L. Ming and H. Song, Neuron {\bf 70}, 687 (2011).
\bibitem{NG6} K. M. Christian, G.-I. Ming, and H. Song, Behav. Brain Res. {\bf 379}, 112346 (2020).
\bibitem{NG7} A. Sahay, D. A. Wilson, and R. Hen, Neuron {\bf 70}, 582 (2011).
\bibitem{NG8} A. Sahay, K. N. Scobie, A. S. Hill, C. M. O’Carroll, M. A. Kheirbek, N. S. Burghardt, A. A. Fenton, A. Dranovsky, and R. Hen,
             Nature {\bf 472}, 466 (2011).
\bibitem{NG9} J. B. Aimone, W. Deng, and F. H. Gage, Neuron {\bf 70}, 589 (2011).
\bibitem{NG10} J. B. Aimone, J. Wiles, and F. H. Gage, Neuron {\bf 61}, 187 (2009).
\bibitem{NG11} C. V. Dieni, R. Panichi, J. B., Aimone, C. T. Kuo, J. I. Wadiche, and L. Overstreet-Wadiche, Nat. Commun. {\bf 7}, 11313 (2016).

\bibitem{SSR} S.-Y. Kim and W. Lim, Cogn. Neurodyn. {\bf 16}, 643 (2022).
\bibitem{PS} S.-Y. Kim and W. Lim, Cogn. Neurodyn. {\bf 16}. 1427 (2022).

\bibitem{BN1} V. Santhakumar, I. Aradi, and I. Soltesz, J. Neurophysiol. {\bf 93}, 437 (2005).
\bibitem{BN2} R. J. Morgan, V. Santhakumar, and I. Soletsz, Prog. Brain Res. {\bf 163}, 639 (2007).

\bibitem{Hilus6} S. Jinde, V. Zsiros, and K. Nakazawa, Front. Neural Circ. {\bf 7}, 14 (2013).

\bibitem{ANA1} M. J. West, L. Slomianka, and H. J. Gundersen, Anat. Rec. {\bf 231}, 482 (1991).

\bibitem{GC-BC1} P. S. Buckmaster, H. J. Wenzel, D. D. Kunkel, and P. A. Schwartzkroin, J. Comp. Neurol. {\bf 366}, 271 (1996).
\bibitem{GC-BC2} P. S. Buckmaster and A. L. Jongen-R{\^{e}}lo, J. Neurosci. {\bf 19}, 9519 (1999).
\bibitem{GC-BC3} P. S. Buckmaster, R. Yamawaki, and G. F. Zhang, J. Comp. Neurol. {\bf 445}, 360 (2002).
\bibitem{GC-BC4} T. Nomura, T. Fukuda, Y. Aika, C. W. Heizmann, P. C. Emson, T. Kobayashi, and T. Kosaka, Brain Res. {\bf 764}, 197 (1997).
\bibitem{GC-BC5} T. Nomura, T. Fukuda, Y. Aika, C. W. Heizmann, P. C. Emson, T. Kobayashi, and T. Kosaka, Brain Res. {\bf 751}, 64 (1997).

\bibitem{ANA3} D. G. Amaral, N. Ishizuka, and B. Claiborne, Prog. Brain Res. {\bf 83}, 1 (1990).
\bibitem{ANA4} B. L. McNaughton, C. A. Barnes, S. J. Y. Mizumori, E. J. Green, and P. E. Sharp, ``Contribution of granule cells to spatial representations in
hippocampal circuits: A puzzle,'' in F. Morrell (ed.). {\it Kindling and Synaptic Plasticity: The Legacy of Graham Goddar}
(Springer-Verlag, Boston, 1991) pp. 110–123.
\bibitem{ANA5} T. Hafting, M. Fyhn, S. Molden, M. B. Moser, and E. I. Moser, Nature {\bf 436}, 801 (2005).
\bibitem{Hilus1} H. E. Scharfman and C. E. Myers, Front. Neural Circ. {\bf 6}, 106 (2013).
\bibitem{Hilus2} H. E. Scharfman, Cell Tissue Res. {\bf 373}, 643 (2018).
\bibitem{Hilus3} J. L\"{u}bke, M. Frotscher, and N. Spruston, J. Neurophysiol. {\bf 79}, 1518 (1998).
\bibitem{Hilus4} D. G. Amaral, H. E. Scharfman, and P. Lavenex, Prog. Brain Res. {\bf 163}, 3 (2007).
\bibitem{Hilus5} S. Jinde, V. Zsiros, Z. Jiang, K. Nakao, J. Pickel, K. Kohno, J. E. Belforte, and K. Nakazawa, Neuron {\bf 76}, 1189 (2012).

\bibitem{Hilus7} A. D. H. Ratzliff, A. L. Howard, V. Santhakumar, I. Osapay, and I. Soltesz, J. Neurosci. {\bf 24}, 2259 (2004).
\bibitem{ANA2} P. S. Buckmaster, A. L. Jongen-R\^{e}, J. Neurosci. {\bf 19}, 9519 (1999).


\bibitem{LIF} W. Gerstner and W. Kistler, {\it Spiking Neuron Models}, (Cambridge University Press, New York, 2002).

\bibitem{NMDA} C. E. Jahr and C. F. Stevens, J. Neurosci. {\bf 10}, 3178 (1990).

\bibitem{SynParm1} T. B. Kneisler and R. Dingledine, Hippocampus {\bf 5}, 151 (1995).
\bibitem{SynParm2} J. R. P. Geiger, J. L\"{u}bke, A. Roth, M. Frotscher, and P. Jonas, Neuron {\bf 18}, 1009 (1997).
\bibitem{SynParm3} M. Bartos, I. Vida, M. Frotscher, J. R. Geiger, and P. Jonas, J. Neurosci. {\bf 21}, 2687 (2001).
\bibitem{SynParm4} C. Schmidt-Hieber, P. Jonas, and J. Bischofberger, J. Neurosci. {\bf 27}, 8430 (2007).
\bibitem{SynParm5} P. Larimer and B. W. Strowbridge, J. Neurosci. {\bf 28}, 12212 (2008).
\bibitem{SynParm6} C. Schmidt-Hieber and J. Bischofberger, J. Neurosci. {\bf 30}, 10233 (2010).
\bibitem{SynParm7} R. Krueppel, S. Remy, and H. Beck, Neuron {\bf 71}, 512 (2011).
\bibitem{SynParm8} P. H. Chiang, P. Y. Wu, T. W. Kuo, Y. C. Liu, C. F. Chan, T. C. Chien, J. K. Cheng, Y. Y. Huang, C. Chiu, Di, and C. C. Lien, J. Neurosci. {\bf 32}, 62 (2012).


\bibitem{Hetero} R. Finnegan and S. Becker, Front. Syst. Neurosci. {\bf 9}, 136 (2015).

\bibitem{AM} S.-Y. Kim and W. Lim, ``Equalization Effect in Interpopulation Spike-Timing-Dependent Plasticity in Two Inhibitory and Excitatory Populations,''
in A. Lintas, P. Enrico, X. Pan, R. Wang, and A. Villa (eds.), {\it Advances in Cognitive Neurodynamics (VII)} (Springer, Singapore, 2021) Ch. 8.
\bibitem{CV} P. Dayan and L. F. Abbott, {\it Theoretical Neuroscience: Computational and Mathematical Modeling of Neural Systems} (MIT press, Cambridge, 2001) Sec. 1.4.

\end{thebibliography}
\end{document}